\documentclass[12pt]{article}
\usepackage{amsfonts,amssymb,latexsym,amscd,amsmath,theorem,mathrsfs,color,bm}
\usepackage[dvipsnames]{xcolor}
\usepackage{tikz}
\usepackage{algorithm} 
\usepackage{authblk}
\usepackage{algpseudocode}
\usepackage[normalem]{ulem}
\usepackage{multirow}
\usepackage{overpic}
\usepackage{tikz}
\usetikzlibrary{shapes.geometric}
\usetikzlibrary{math}
\usetikzlibrary {arrows.meta}
  
\graphicspath{{./images/}}
\DeclareMathOperator{\grad}{grad}

\newcommand{\ee}{\mathscr{E}}

\newcommand{\EE}{\mathscr{E}} 
\newcommand{\CC}{\mathscr{C}}

\newcommand{\R}{{\mathbb{R}}}
\newcommand{\Z}{{\mathbb{Z}}}

\newcommand{\C}{{\mathbb{C}}}
\newcommand{\I}{{\mathbb{I}}}
\newcommand{\DD}{{\mathscr{D}}}

\newcommand{\beq}{\begin{equation}}
\newcommand{\eeq}{\end{equation}}
\newcommand{\bea}{\begin{eqnarray}}
\newcommand{\eea}{\end{eqnarray}}
\newcommand{\ben}{\begin{eqnarray*}}
\newcommand{\een}{\end{eqnarray*}}
\newcommand{\bem}{\begin{enumerate}}
\newcommand{\eem}{\end{enumerate}}

\newcommand{\ra}{\rightarrow}

\newcommand{\ds}{\displaystyle}
\newcommand{\cd}{\partial}
\newcommand{\wt}{\widetilde}

\newcommand{\ip}[1]{\langle #1 \rangle}
\newcommand{\ignore}[1]{}

\renewcommand{\star}{*}

\newcommand{\hh}{\mathscr{H}}

\newcommand{\vecs}[1]{{\bm #1}}

\newcommand{\ol}{\overline}

\newcommand{\avec}{\mbox{\boldmath{$a$}}}
\newcommand{\nvec}{\mbox{\boldmath{$n$}}}

\newcommand{\mvec}{\mbox{\boldmath{$m$}}}

\newcommand{\xvec}{\mbox{\boldmath{$x$}}}
\newcommand{\svec}{\mbox{\boldmath{$s$}}}
\newcommand{\uvec}{{\mbox{\boldmath{$u$}}}}
\newcommand{\vvec}{\mbox{\boldmath{$v$}}}
\newcommand{\dvec}{\mbox{\boldmath{$d$}}}
\newcommand{\dvecs}{{\bm d}}

\newcommand{\evec}{{\mbox{\boldmath{$e$}}}}
\newcommand{\Hvec}{\mbox{\bm{$H$}}}
\newcommand{\Hvecs}{\bm{H}}

\newcommand{\Nvec}{\mbox{\boldmath{$N$}}}

\newcommand{\Xvec}{\mbox{\boldmath{$X$}}}
\newcommand{\Svec}{\mbox{\boldmath{$S$}}}
\newcommand{\zerovec}{\mbox{\boldmath{$0$}}}

\newcommand{\tr}{{\rm tr}\, }

\newcommand{\eps}{\varepsilon}

\theoremstyle{plain}

{\theorembodyfont{\rmfamily}

}

\newcommand{\news}{\setcounter{equation}{0}}

\newcommand{\figref}{figure \ref}

\newcommand{\SP}{SP }

\newcommand{\tikzcircle}[2][red]{\tikz[baseline=-0.5ex]\draw[#1,radius=#2,fill=#1] (0,0) circle ;}%

\definecolor{TW-color}{RGB}{100,0,100}
\definecolor{Error-color}{RGB}{250,0,0}

\begin{document}
 
\title{Towards a universal phase diagram of \\ planar chiral magnets}
\author[1]{Bernd Schroers\thanks{E-mail: {\tt B.Schroers@ed.ac.uk}}}
\author[2]{Martin Speight\thanks{E-mail: {\tt j.m.speight@leeds.ac.uk}}}
\author[1]{Thomas Winyard\thanks{E-mail: {\tt twinyard@ed.ac.uk}}}
\affil[1]{Maxwell Institute of Mathematical Sciences and School of Mathematics\\
University of Edinburgh, Edinburgh\\
EH9 3FD, United Kingdom}
\affil[2]{School of Mathematics, University of Leeds\\
Leeds LS2 9JT, England}

\renewcommand\Affilfont{\itshape\small}

\newcommand{\dmi}{b}

\maketitle

\begin{abstract}
\noindent In planar chiral magnets, the competition of the positive definite Heisenberg exchange  and Zeeman energies with the indefinite  Dzyaloshinskii–Moriya interaction (DMI) energy allows for the possibility  of negative energy ground states,  and leads to an intricate dependence of the ground states on the parameters of the theory. In this paper, we consider arbitrary spiralization tensors for the DMI interaction and arbitrary directions for the external magnetic field,  and study the nature of the  ground states in this parameter space, using a combination of analytical and numerical methods. Classifying ground states  by their symmetry into ferromagnetic (invariant under under arbitrary translations in the plane), spiral  (invariant under arbitrary translations in one direction) and  skyrmion lattice ground states (invariant under a two dimensional lattice group),  we give a complete  description of the phase diagram of this class of theories. 
\end{abstract}

\setcounter{section}{-1}

\section{Introduction}
Most materials have lattice structures in their solid state. This is relatively straightforward to understand in quantum-mechanical descriptions of solids because periodic arrangements of atoms  minimize the total interaction energy of a large number of atoms. In nonlinear field theories of matter, however, the nature of the ground state and its dependence on order parameters in the model is often one of the most intricate and interesting aspects of the theory.

This is particularly true if the field theory permits topological solitons - particle-like solutions of the field equations with a localized energy density which are stable for topological reasons. Such solitons have an associated topological invariant which one can typically think of as the number of particles in a configuration. Minimal energy configurations need to be determined separately in each topological sector of the theory. In analogy with atomic models of solids, one would then expect lattices to appear in topological sectors containing many particles. This is indeed the case, for example, in the Skyrme model of nuclear particles, where solitons - called Skyrmions - are models for nucleons \cite{manton2004topological}. Skyrmions repel at small separation, but attract at large separation for suitable relative orientations. As a result, the interaction potential has a minimum at a finite separation, and lattices can, and indeed do, form by a mechanism not dissimilar from the one leading to atomic lattices in quantum mechanics.

There is a further and rather different mechanism by which lattices of solitons form in field theories, which is common in theories of condensed matter systems, subject to an externally applied magnetic field. The seminal example is the Abrikosov lattice in Ginzburg-Landau models of vortices \cite{abrikosov1957magnetic}, and later the theory of chiral magnetic skyrmions \cite{bogdanov1989thermodynamically} which we study in this paper. In such theories, there is typically a homogeneous ground state for a strong external field (normal/ferromagnetic state). However, as one reduces the external field, the system can lower its total energy by creating topological defects which have negative energy relative to the homogeneous ground state. If such defects attract each other, they can form clusters of arbitrarily negative energy. However, if they repel, one expects there to be an equilibrium where topological defects form an infinite arrangement with finite separations between them. If that arrangement is doubly periodic, the ground state is a lattice of solitons. If it is translation invariant in one direction but periodic in another, then it is a lattice of line defects.

In this paper, we carry out a systematic study of ground states in chiral magnets in the plane.  The model allows for topological solitons, called chiral magnetic skyrmions, which  were predicted theoretically by Bogdanov and Yablonskii \cite{bogdanov1989thermodynamically}, and later observed experimentally – in the form of a spatially periodic lattice – by M\"uhlbauer et al \cite{Muhlbauer2009}.  They are of potential interest  for next-generation and low-energy data storage technology \cite{Fert2013Skyrmions}, and have been the focus of intense theoretical and experimental study for several years \cite{fert2017magnetic,tokura2020magnetic}.

The basic field (or order parameter) of the model is the magnetization vector, mathematically a smooth map $\mvec:\R^2\rightarrow S^2$. The energy functional  contains the usual Heisenberg exchange interaction, the Dzyaloshinskii–Moriya interaction (DMI) and the Zeeman interaction with an externally applied magnetic field. As we explain in the paper, it is useful to think of the parameters in the  DMI term (often called spiralization tensor) as those of an ellipse in three dimensional space. Using the symmetries of the theory, we can assume that this ellipse lies in the $x_1x_2$-plane, and has a major axis along the $x_1$-direction and of length one. The length of the minor axis is an irreducible parameter of the DMI, varying from zero for  so-called rank one materials to one for the much-studied axisymmetric DMI term. Having used the symmetries to bring the DMI into a standard form, we have to allow for arbitrary magnitude and direction of the magnetic field, so that our model has a four dimensional phase space. General anisotropy terms would add further parameters, but we do not consider them in this paper.

For sufficiently strong external magnetic fields, the energy is dominated by the Zeeman term: the magnetization vector aligns with the magnetic field everywhere, leading to homogeneous ferromagnetic vacuum state of total energy zero. However, for vanishing or weak magnetic fields, the DMI term determines the ground state: helical configurations of the magnetization vector, which have  negative DMI energy density, may then have total energy per unit area which is also negative. Such configurations  form the new ground state of the theory. For intermediate magnitudes of the external magnetic field, there is a subtle interplay of all three energy terms, which, for certain regions of phase space, leads to the minimization of the energy per unit area through magnetic skyrmions.  These are topological solitons of topological degree minus one, and with negative total energy.  Their repulsive interaction \cite{barton2023stability} prevents clustering, but allows for the formation of lattices. 

Ground states of planar chiral magnets have been studied for particular parameter choices in, for example, \cite{bogdanov1994thermodynamically} and more systematically for general DMI terms and with the magnetic field orthogonal to the plane in \cite{Gungordu2016SkyrmionStability}.
In this paper, we carry out a  comprehensive study of  the ground state of the theory  for general  DMI  parameters and arbitrary strength and direction of the  external magnetic fields. In this way we arrive at a phase diagram of planar chiral magnets which divides the four dimensional parameter space into ferromagnetic, spiral and skyrmion lattice phases. 
Our numerical method for finding spiral ground states and skyrmion lattices is based on techniques which two of the authors developed for vortex lattices in superconductors initially in \cite{speight2023symmetries} and generalized in \cite{SpeightWinyard2025}. Applied to skyrmion lattices, this involves minimizing the energy per unit area of the lattice with respect to the magnetization field and, crucially, also the period lattice. This is closely analogous to the problems studied in the context of multicomponent superconductors in \cite{speight2023magnetic,wang2025observation} and can be solved by an efficient gradient descent method called arrested Newton flow.

The paper is organized as follows. In Section 1, we define the model of planar chiral magnets in terms of its energy functional, explain the geometrical representation of the DMI  parameters in terms of an ellipse, and use symmetries to bring the energy into a standard form which depends on four parameters. In Section 2, we study configurations which are translation invariant in one direction and periodic in another and which, following the conventions in the literature, we call spiral. We define the spiral domain $\hh_b$  as the set of  magnetic fields for which the total energy is negative when evaluated on a spiral configuration, and use three  ans\"atze for such configurations in terms of helical fields, periodic arrangements of stripes, and conical fields  to determine bounds on $\hh_b$. Section 3 contains a careful exposition of our numerical method for reliably determining lattice ground states. The results of this scheme are presented in Section 4: we illustrate our insights into the four dimensional phase diagram in terms of various two dimensional phase plots before turning to a final discussion of our results in the concluding Section 5.

\section{The model}\news

Consider the following free energy,
\beq
E = \int_{\R^2} \left\{\frac{1}{2} |d \mvec|^2 + D_{ai} \eps_{abc} m_b \frac{\partial m_c}{\partial x_i} + V(\mvec)\right\}d^2x
=E_{exch}+E_{DMI}+E_{pot}
\label{eq:E}
\eeq
where $\mvec : \mathbb{R}^2 \rightarrow S^2$ is the magnetization and $D$ is the spiralization tensor, a constant $3\times 2$ real matrix that determines the form of the DMI term. Throughout the paper we will use letters at the beginning of the alphabet for target space indices $a \in \{1,2,3\}$ and in the middle of the alphabet for (thin film) spatial indices $i \in \{1,2\}$. The potential $V$ is a smooth function $S^2\ra\R$ which, without loss of generality, we assume attains a minimum value of $0$. Many potentials are phenomenolgically interesting, but we will restrict attention to the numerous materials where anisotropy is negligible, so that our potential is simply the Zeeman term associated with an applied external magnetic field $\Hvec$,
\beq
V(\mvec) = |\Hvec|- \Hvec \cdot \mvec.
\label{eq:V}
\eeq
This is minimized when $\mvec=\hat\Hvec:=\Hvec/|\Hvec|$, and the Euler-Lagrange equation for $E$ always admits the homogeneous solution $\mvec(x) = \hat\Hvec$ with energy $E=0$. This is the so-called ferromagnetic state.  

Note that the homogeneous solution is not necessarily the global minimizer of $E$, as the DMI term can be negative. Indeed, any finite energy critical point of $E$ must have $E_{DMI}\leq 0$ by the usual Derrick scaling argument \cite{der}. Assume $\mvec:\R^2\ra S^2$ is a critical point of $E$ and denote by $\mvec_\lambda$ the one-parameter variation of $\mvec$ by spatial dilation, $\mvec_\lambda(\xvec)=\mvec(\lambda\xvec)$. A simple rescaling of the spatial coordinate reveals that
\beq
E(\mvec_\lambda)=E_{exch}(\mvec)+\frac{1}{\lambda}E_{DMI}(\mvec)+\frac{1}{\lambda^2}E_{pot}(\mvec).
\eeq
Since $E$ is, by assumption, critical with respect to all smooth variations of $\mvec$, it is critical with respect to the variation $\mvec_\lambda$, that is
\beq
\frac{d\:}{d\lambda}\bigg|_{\lambda=1}E(\mvec_\lambda)=-E_{DMI}(\mvec)-2E_{pot}(\mvec)=0.
\eeq
Hence all finite energy static solutions satisfy the virial constraint
\beq\label{virial}
E_{DMI}(\mvec)=-2E_{pot}(\mvec)
\eeq
implying that $E_{DMI}(\mvec)\leq 0$ (with equality if and only if $\mvec$ is the ferromagnetic state). 

\begin{figure}
\begin{overpic}[width=0.48\linewidth]{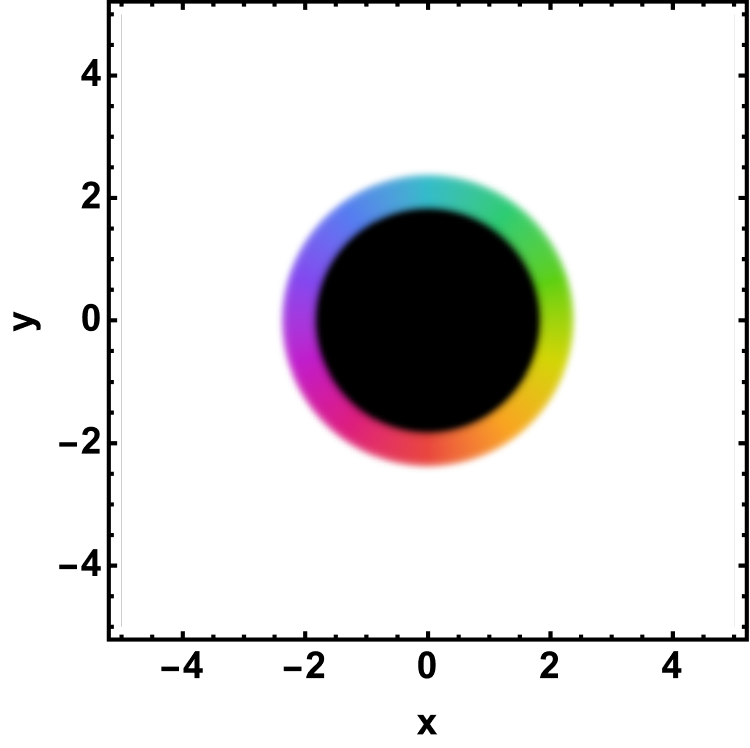}
\end{overpic}
\begin{overpic}[width=0.48\linewidth]{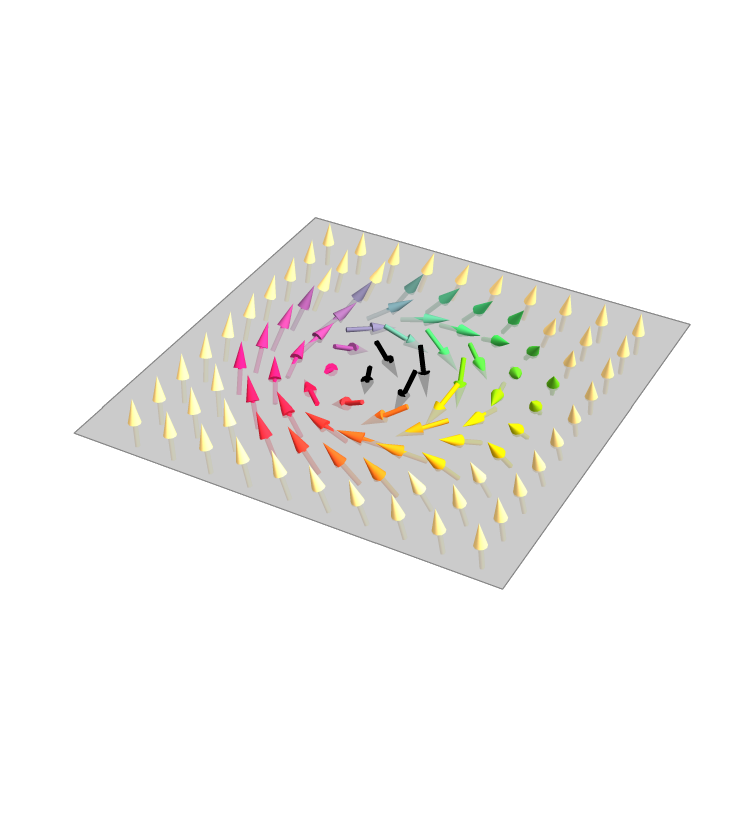}
\put(31,30){$\boldsymbol{x}$}
\put(20,65){$\boldsymbol{y}$}
\end{overpic}
\label{fig:skyrmion}
\caption{The magnetization field $\mvec$ for a single skyrmion ($Q = -1$), for the model with standard isotropic DMI term and applied magnetic field $\Hvec=(0,0,1)$. Left and right panels depict the same field. The left panel records the orientation of $\mvec$ by assigning a colour to each point on $S^2$, while the right panel shows $\mvec$ using oriented arrows in $\R^3$ coloured using the same rule. So black denotes $(0,0,-1)$ and white denotes $(0,0,1)$, for example. We will use the colouring scheme of the left panel throughout the paper.}
\end{figure}

Fields which tend to the ferromagnetic state as $|\xvec|\ra\infty$ are classified by their topological degree
\beq
Q(\mvec)=\frac{1}{4\pi}\int_{\R^2}\mvec\cdot(\cd_1\mvec\times\cd_2\mvec) \in \Z
\eeq
which may be interpreted as the number of times $\mvec$ wraps $\R^2\cup\{\infty\}\equiv\ S^2$ around $S^2$. Minimizers of $E$ among fields of degree $Q=-1$ ($Q=1$) are conventionally called skyrmions (antiskyrmions). A numerically generated skyrmion for the model with $D_{ai}=\delta_{ai}$ and $\Hvec=(0,0,1)$ is shown in \figref{fig:skyrmion}. The focus of this paper is skyrmion {\em lattices} (and other spatially periodic solutions), not isolated skyrmions such as this. We depict it here mainly to illustrate the colouring scheme we will use consistently to represent $\mvec$.

\subsection{Reduction of DMI parameters}

In this section, we show how field and spatial coordinate redefinitions can be used to reduce the general DMI term, containing 6 real parameters $D_{ai}$, to a one-parameter family. 

\begin{figure}[h]
    \centering
         \includegraphics[trim={10.5cm 13cm 16cm 11cm},clip,width=0.7\textwidth]{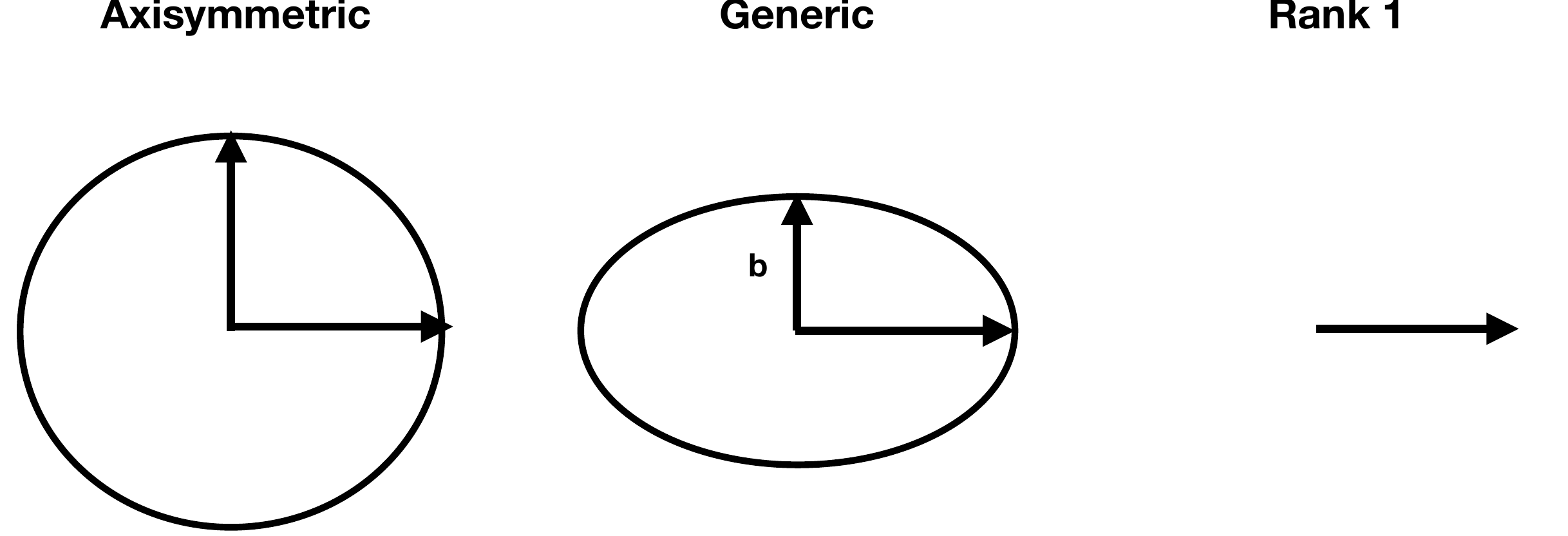}
\caption{DMI terms classified in terms of ellipses with major axis 1, minor axis $b$ and  eccentricity $\varepsilon =\sqrt{1-b^2} $}
\label{fig:DMIellipse} 
\end{figure}

Let $\dvec_1,\dvec_2\in\R^3$ denote the columns of the spiralization tensor. Then the DMI energy is
\beq
{E}_{DMI}=\int_{\R^2}\sum_{i=1,2}\dvec_i\cdot(\mvec\times\cd_i\mvec)d^2x.
\eeq
It is geometrically natural to think of the spiralization tensor in  terms of the ellipse traced out by
\beq 
\dvec(\theta)= \dvec_1 \cos\theta + \dvec_2\sin   \theta, \quad \theta \in [0,2\pi),
\label{DMIellipse}
\eeq
see \figref{fig:DMIellipse}.
This picture, which includes the degenerate cases of the circle ($\dvec_1$ and $\dvec_2$ orthogonal and of equal magnitude) and the line ($\dvec_1$ and $\dvec_2$ linearly dependent but not both zero) is natural because it captures the symmetry of the DMI energy. It also immediately suggests a route to standardizing the form of the DMI interaction by rotating and scaling the vectors $\dvec_1$ and $\dvec_2$ into a standard form in the 12-plane of $\R^3$. We explain this carefully below, but to motivate the steps we take, we note that the lengths and directions of the  major and minor axes  of the ellipse defined by \eqref{DMIellipse} are the eigenvalues and eigenvectors of the positive definite and symmetric  Gram matrix $G_{ij}= \dvec_i\cdot \dvec_j$. This  matrix can be brought into a standard diagonal form  with  ordered eigenvalues 
\beq 
\lambda^\pm=\frac{|\dvec_1|^2+|\dvec_2|^2}{2}\pm \frac 12 
\sqrt{(|\dvec_1|^2-|\dvec_2|^2)^2+(2\dvec_1\cdot\dvec_2)^2}.
\label{Geigen}
\eeq
by conjugation with a suitable $SO(2)$ matrix $R$
which rotates $\dvec_i \mapsto \wt\dvec_i=\sum_jR_{ij}\dvec_j$. 
In particular, we therefore have $\wt{G}_{12}=\wt\dvec_1\cdot\wt\dvec_2=0$ and $\wt{G}_{11}=|\wt\dvec_1|^2\geq \wt{G}_{22}=|\wt\dvec_2|^2$  in this basis.

We can use these considerations to simplify the energy expressions by 
combining the rotation of the DMI vectors with  rotated spatial coordinates $\wt{x}_i=\sum_{j} R_{ij}x_j$.  The energy is now
\beq
{E}_{DMI}=\int_{\R^2}\sum_{i=1,2}\wt\dvec_i\cdot(\mvec\times\frac{\cd\mvec}{\cd \wt{x}_i})d^2\wt{x},
\eeq
We also define rescaled coordinates, $\ol{x}_i=|\wt{d}_1|\wt x_i$. Then
\beq
E=\int_{\R^2}\left\{\frac12\sum_i\left|\frac{\cd\mvec}{\cd \ol{x}_i}\right|^2
+\sum_i \ol{\dvec}_i\cdot\left(\mvec\times\frac{\cd\mvec}{\cd \ol{x}_i}\right)+|\ol\Hvec|-\mvec\cdot\ol\Hvec\right\}d^2\ol{x}
\eeq
where $\ol\dvec_i=\wt\dvec_i/|\wt\dvec_1|$ and $\ol\Hvec=\Hvec/|\wt{\dvec}_1|^2$.
Note that $|\ol\dvec_1|=1$ by construction. 

We now observe that there exists $S\in SO(3)$ (unique, unless $\ol\dvec_2=\zerovec$) such that
\beq
\dvec_1':=S\ol\dvec_1=(1,0,0),\qquad
\dvec_2':=S\ol\dvec_2=(0,b,0),
\eeq
where $b\in[0,1]$. If we rotate $\mvec$ by the same matrix, $\mvec':= S\mvec$, we see that
\beq
E=\int_{\R^2}\left\{\frac12\sum_i\left|\frac{\cd\mvec'}{\cd \ol{x}_i}\right|^2
+\sum_i {\dvec}_i'\cdot\left(\mvec'\times\frac{\cd\mvec'}{\cd \ol{x}_i}\right)+|\Hvec'|-\mvec'\cdot\Hvec'\right\}d^2\ol{x}
\eeq
where $\Hvec'=S\ol\Hvec$. This is the final reduced form of our energy functional, and it depends on the single parameter $b\in[0,1]$, and the (rescaled, rotated) applied field $\Hvec'$. It is important to note that, in achieving this reduction, we have broken the correlation between the axis directions in the sample plane and in the field space. In particular, when we speak of the orthogonal (or parallel) component of $\Hvec$ or $\mvec$, we mean the component orthogonal (or parallel) to the plane spanned by the spiralization vectors $\dvec_1,\dvec_2$, and this, in general, differs from the sample plane (spanned by $(1,0,0)$, $(0,1,0)$). 

A quick way to identify which DMI term in our one-parameter family
corresponds to an unreduced pair $(\dvec_1,\dvec_2)$ is to note that
\beq
b^2=\frac{|\wt\dvec_2|^2}{|\wt\dvec_1|^2}=\frac{\lambda^-}{\lambda^+},
\eeq
where $\lambda^\pm$ are the eigenvalues \eqref{Geigen}.
So if $\dvec_1,\dvec_2$ are orthogonal and of equal length, $b=1$ (the standard case), whereas if $\dvec_1,\dvec_2$ are linearly dependent, $b=0$ (the rank 1 case).

From now on, we drop all decoration from $\mvec$, $\Hvec$, $(x_1,x_2)$ and $\dvec_i$, the preceding rotation, rescaling, rotation procedure being implicitly assumed. When we wish to emphasize the specific choice of DMI parameter $b\in[0,1]$ and applied field $\Hvec\in\R^3$ under consideration, we will denote the energy functional $E_{b,\Hvecs}$, so
\beq
E_{b,\Hvecs}(\mvec)=\int_{\R^2}\left\{\frac12(|\cd_1\mvec|^2+|\cd_2\mvec|^2)+
(\mvec\times\cd_1\mvec)_1+b(\mvec\times\cd_2\mvec)_2+|\Hvec|-\Hvec\cdot\mvec\right\}d^2x.
\label{eq:EbH}
\eeq
Note that the models with $b=1$ and $b=0$ each enjoy a rotation symmetry, since for all fields $\mvec:\R^2\ra S^2$, all applied fields $\Hvec\in\R^3$ and rotations
$R_3$, $R_1$ about the $3$ and $1$ axes respectively,
\bea
E_{1,\Hvecs}(R_3\circ \mvec\circ R_3^{-1})&=&E_{1,R_3^{-1}\Hvecs}(\mvec), \label{1sym}\\
E_{0,\Hvecs}(R_1\circ\mvec)&=&E_{0,R_1^{-1}\Hvecs}(\mvec).\label{0sym}
\eea

\subsection{Rank 1 materials}\label{rank1}

Consider the case where $\dvec_1$ and $\dvec_2$ are linearly dependent, hence $b=0$ and, in the coordinate system derived above, $\dvec_1=(1,0,0)$, $\dvec_2=\zerovec$. Then
\beq
E_\text{DMI} = \int_{\mathbb{R}^2} \dvec_1 \cdot \mvec \times \partial_1 \mvec.
\eeq
We now demonstrate that this model can have no spatially localized static solutions of negative total energy. Hence skyrmion crystals, if they exist, can only be metastable in these systems, never the ground state.  

To see this, let us split the exchange energy into two terms,
\begin{align}
E_{exch} &= E^{(1)}_{exch} + E^{(2)}_{exch},&
E_{exch}^{(i)} &= \frac{1}{2}\int_{\mathbb{R}^2} \left|\partial_i \mvec\right|^2,
\end{align}
and consider the anisotropic scaling $\mvec_\lambda(\xvec):=\mvec(\lambda x_1,x_2)$ of a static solution $\mvec$, whose energy is,
\beq
E(\mvec_\lambda) = \lambda E_{exch}^{(1)}(\mvec) + \frac{1}{\lambda} E_{exch}^{(2)}(\mvec) + E_\text{DMI}(\mvec) + \frac{1}{\lambda} E_{pot}(\mvec).
\eeq
As in Derrick's argument, this function of $\lambda$ must have a critical point at $\lambda=1$, so
\beq\label{virial2}
\left.\frac{dE(\mvec_\lambda)}{d\lambda} \right|_{\lambda = 1} = E_{exch}^{(1)} - E_{exch}^{(2)} - E_{pot} = 0.
\eeq
Combining \eqref{virial} with \eqref{virial2} we see that
\beq
E = E_{exch}^{(1)} + E_{exch}^{(2)} + E_\text{DMI} + E_{pot} = 
 E_{exch}^{(1)} + E_{exch}^{(2)}-E_{pot}=2E_{exch}^{(2)}\geq 0.
\eeq 
Hence, every spatially localized (and hence finite energy) solution of the rank 1 model has positive energy.

One should note that the argument above applies only to spatially localized fields. As we will see shortly, the ground state of the rank 1 model with sufficiently small $|\Hvec|$ is not the ferromagnetic state since there exist one dimensional spatially periodic fields (spiral phases) of negative average energy density. Such fields evade the argument above since they have infinite $E$.

\section{The spiral phase}\label{SP}\news

\subsection{The spiral domain}

Recall that the ferromagnetic state, $\mvec(\xvec)=\hat\Hvec:=\Hvec/|\Hvec|$ constant, has zero energy, but may fail to be the ground state of the system, since the DMI energy can be negative. In this section we will consider {\em spiral phase} solutions of the system. By definition, these are solutions of the Euler-Lagrange equation for $E$ which are translation invariant in some direction in the sample plane and vary periodically (spatially) in the orthogonal direction. If these have negative energy, they are energetically preferred over the FM state, and are candidates for the system's ground state (though, as we will see later, more complex configurations with less symmetry may have even lower energy). For a given DMI parameter $b\in[0,1]$, we  denote by $\hh_b\subset\R^3$ the {\em spiral domain} of the system, that is, the set of applied magnetic field values $\Hvec$ for which $\ds{E_{b,{\Hvecs}}}$ \eqref{eq:EbH} has a negative energy spiral phase solution. Our aim in this section is to understand $\hh_b$ as completely possible, by means of rigorous analytic bounds and careful numerics.

It will be useful to distinguish between spiral phase solutions and the weaker notion of a {\em spiral phase} of the system, which we define to be any field $\mvec:\R^2\ra S^2$ which is translation invariant in some direction in $\R^2$, and varies periodically in the direction orthogonal to invariance. Explicitly, let $\nvec\in S^1\subset\R^2$ be a unit vector in the $(x_1,x_2)$ plane, and assume that $\mvec$ is invariant under translations orthogonal to $\nvec$, so 
\beq
\mvec(\xvec)=\svec(\nvec\cdot\xvec)
\eeq
for some periodic map $\svec:\R\ra S^2$ of period $T>0$. This is a spiral phase {\em solution} if and only if $\svec$ is a critical point of the dimensionally reduced energy functional
\beq
E = \int \left\{\frac{1}{2} |\dot{\svec}(t)|^2 + \dvec(\nvec)\cdot\left( \svec(t)\times\dot{\svec}(t)\right) + |\Hvec| - \Hvec\cdot \svec(t)\right\}\, dt,
\label{eq:E1}
\eeq
where $t = \nvec\cdot\xvec$ and we have defined,
\beq
\dvec(\nvec) = n_1 \dvec_1 + n_2 \dvec_2,
\eeq
a vector on the DMI ellipse.
As an interesting aside, the functional \eqref{eq:E1} is the Lagrangian for a charged particle moving on a sphere in the presence of a magnetic field $\dvec(\nvec)$ and gravitational field $-\Hvec$, so critical points of $E$ may be reinterpreted as time-dependent trajectories of such a particle. We emphasize, however, that $E$ is the energy functional of our system, not a Lagrangian, and $t$ represents a spatial variable, measuring distance in the sample plane along the direction orthogonal to $\nvec$.  We require periodic solutions of this problem, of period $T>0$ (so $\svec(t+T)\equiv\svec(t)$), and vary $\svec$ to minimize the average energy density
\beq
\ip{\ee} = \frac1T\int_0^T \left\{\frac{1}{2} |\dot{\svec}(t)|^2 + \dvec(\nvec)\cdot\left( \svec(t)\times\dot{\svec}(t)\right) + |\Hvec| - \Hvec\cdot \svec(t)\right\}\, dt,
\label{eq:E1average}
\eeq
subject to periodic boundary conditions. 

By a standard application of the direct method of the calculus of variations in the Hilbert space $H^1(\R/T\Z,S^2)$, such a critical point exists, for any fixed $T>0$ and $\nvec\in S^1$, attaining the infimum of $\ip\EE$ on this space. Hence, to show that $\Hvec\in\hh_b$ it suffices to construct any spiral phase for $E_{b,\Hvecs}$ of negative average energy density: the spiral phase solution with the same $\nvec$ and $T$ has $\ip{\ee}$ no greater than this, and hence also has $\ip{\ee}<0$.

Generalizing the discussion of helical phases in \cite{Melcher2014chiral}, we re-write the average energy density \eqref{eq:E1average} by completing the square: 
\beq
\ip{\ee} = \frac1T\int_0^T \left\{\frac{1}{2} |\dot{\svec}(t)+ \dvec \times \svec(t)|^2  -\frac 12 (|\dvec|^2 -(\dvec\cdot\svec)^2) + |\Hvec| - \Hvec\cdot \svec(t)\right\}\, dt,
\label{eq:E2average}
\eeq
Helical configurations of the form 
\beq
\svec(t) = R_{\hat{\dvecs}}(-t)\svec_0,
\eeq
where  $R_{\hat{\dvec}}(-t)$  denotes a rotation about the axis $\hat\dvec$ by an angle $-t$,  satisfy  $\dot{\svec}(t)+ \dvec \times \svec(t)=0$. If we choose $\svec_0$ orthogonal to $\dvec$, we obtain $\svec(t)=  \svec_0 \cos t- \hat{\dvec} \times \svec_0 \sin t $, and calculate
\beq
\ip{\ee} = - \frac 12 |\dvec|^2 + |\Hvec|. 
\eeq
Choosing $\nvec = (1,0)$, we conclude that, for $|\Hvec|< \frac 12  $, there must always exist a spiral phase of negative energy. Hence, for all $b\in[0,1]$, $B_{1/2}(0)\subseteq\hh_b$ where $B_\rho(\avec)$ denotes the open ball in $\R^3$ of radius $\rho$ and centre $\avec$.

On the other hand, one can easily prove that if $|\Hvec|\geq 1$ there can be no spiral phase of negative energy. Let $\svec_0=\svec-\hat{\Hvec}:[0,T]\ra\R^3$.
Then,
using $\ip{\cdot,\cdot}$ and $\|\cdot\|$ to denote the $L^2$ inner product and $L^2$ norm, respectively, on functions $[0,T]\ra\R^3$, we have the identities
\bea
E&=&\frac12\|\dot\svec_0\|^2-\ip{(\svec_0+\hat\Hvec)\times\dvec,\dot\svec_0}+
\frac{|\Hvec|}{2}\|\svec_0\|^2 \nonumber \\
&=&\frac12\|\dot\svec_0\|^2-\ip{\svec_0\times\dvec,\dot\svec_0}
-(\hat\Hvec\times\dvec)\cdot\int_{[0,T]}\dot\svec_0+
\frac{|\Hvec|}{2}\|\svec_0\|^2\nonumber \\
&\geq&\frac12\|\dot\svec_0\|^2-\|\svec_0\times\dvec\|\|\dot\svec_0\|-0+
\frac{|\Hvec|}{2}\|\svec_0\|^2\nonumber \\
&\geq&\frac12\|\dot\svec_0\|^2-\|\svec_0\|\|\dot\svec_0\|+
\frac{|\Hvec|}{2}\|\svec_0\|^2\nonumber \\
&\geq&\frac12\left(\|\dot\svec_0\|-\|\svec_0\|\right)^2-\frac12\|\svec_0\|^2+
\frac{|\Hvec|}{2}\|\svec_0\|^2\nonumber \\
&\geq&\frac{(|\Hvec|-1)}{2}\|\svec_0\|^2\geq 0\label{quhama}
\eea
where we have used the periodicity of $\svec_0$, $|\dvec|\leq 1$ and the Cauchy-Schwarz inequality.  Hence
$\hh_b\subseteq B_1(0)$.

Several other properties of $\hh_b$ follow easily from its definition. It is clear that $\hh_b$ is {\em open}, since if $\mvec$ is a negative energy spiral phase for field $\Hvec_0$, the same field is a negative energy spiral phase for all $\Hvec$ sufficiently close to $\Hvec_0$.  
By similar reasoning, we see that $\hh_b$ is {\em star-shaped} with centre $\zerovec$, that is, if $\Hvec\in\hh_b$, then for all $\tau\in[0,1]$, $\tau\Hvec\in\hh_b$. To see this note that
if $\mvec$ is a negative energy spiral phase for field $\Hvec$, the same field is a negative energy spiral phase for $\tau\Hvec$ for all $\tau\in[0,1]$ since this change reduces $E_{pot}$ and leaves $E_{exch}$, $E_{DMI}$ unchanged. 
For all $b\in[0,1]$, $\hh_b$ has several reflection symmetries:
if $(H_1,H_2,H_3)\in\hh_b$ then so are $(-H_1,H_2,H_3)$, $(H_1,-H_2,H_3)$ and $(H_1,H_2,-H_3)$. The extreme cases $b=0$ and $b=1$ have enlarged symmetry: $\hh_1$ is a volume of revolution about the $H_3$ axis, by the symmetry \eqref{1sym}, while
$\hh_0$ is a volume of revolution about the $H_1$ axis, by  the symmetry \eqref{0sym}.

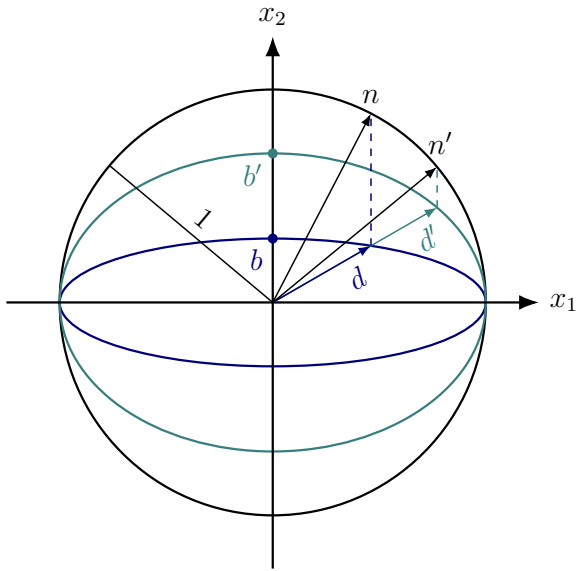
\begin{figure}
\begin{center}
\resizebox{0.5\textwidth}{!}{
\definecolor{col1}{RGB}{0, 0, 123}
\definecolor{col2}{RGB}{55, 126, 126}

\begin{tikzpicture}
\def \scale {3.0};
\def \b {0.3*\scale};
\def \bd {0.7*\scale};
\def \dAngle {30};
\def \dLength {(\scale*sqrt(1.0/(cos(\dAngle)^2 + ((sin(\dAngle)^2)/((\b/\scale)^2)))))};
\def \dDashLength {(\scale*sqrt(1.0/(cos(\dAngle)^2 + ((sin(\dAngle)^2)/((\bd/\scale)^2)))))};
\def \ellipseWidth {0.3*\scale};
\def \axisWidth {0.3*\scale};
\def \axisLength {1.25*\scale};
\def \dWidth {0.2*\scale}
\def \radiusAngle {140}
\coordinate (dp) at ({\dLength*cos(\dAngle)}, {\dLength*sin(\dAngle)});
\coordinate (ddp) at ({\dDashLength*cos(\dAngle)}, {\dDashLength*sin(\dAngle)});
\def \dCircProjx {\dLength*cos(\dAngle)};
\def \dDashCircProjx {\dDashLength*cos(\dAngle)};
\coordinate (dCircProj) at ({\dCircProjx}, {sqrt(((\scale)^2) - ((\dCircProjx)^2))});
\coordinate (dDashCircProj) at ({\dDashCircProjx}, {sqrt(((\scale)^2) - ((\dDashCircProjx)^2))});
\draw[line width = \ellipseWidth] (0, 0) ellipse ({\scale} and {\scale});
\draw[line width = \ellipseWidth, color=col1] (0, 0) ellipse ({\scale} and \b);
\draw[line width = \ellipseWidth, color=col2] (0, 0) ellipse ({\scale} and \bd);
\draw[line width = \axisWidth, -{Latex[length=8]}] (-\axisLength, 0) -- (\axisLength, 0) node[right]{$x_{1}$};
\draw[line width = \axisWidth, -{Latex[length=8]}] (0, -\axisLength) -- (0, \axisLength) node[above]{$x_{2}$};
\fill[color=col2] (0,\bd) circle[radius=2pt] node[below left] {$b'$};
\fill[color=col1] (0,\b) circle[radius=2pt] node[below left] {$b$};
\draw[line width = \dWidth, -{Latex[length=5]},color=col2] (0, 0) -- (ddp) node[pos=0.88,below,sloped]{$d'$};
\draw[line width = \dWidth, -{Latex[length=5]}, color=col1] (0, 0) -- (dp) node[pos=0.75,below,sloped]{$d$};
\draw[line width = \dWidth, -{Latex[length=5]}] (0, 0) -- (dCircProj) node[above]{$n$};
\draw[line width = \dWidth,  -{Latex[length=5]}] (0, 0) -- (dDashCircProj) node[pos=1.02,above]{$n'$};
\draw[line width = \dWidth, dashed, color=col1] (dp) -- (dCircProj);
\draw[line width = \dWidth, dashed, color=col2] (ddp) -- (dDashCircProj);
\draw[line width = \dWidth] (0,0) -- ({\scale*cos(\radiusAngle)},{\scale*sin(\radiusAngle)}) node[midway,above,sloped]{$1$};
\end{tikzpicture}
}
\end{center}
\caption{The nesting of the ellipses implies a nesting property of the spiral domain: $\hh_b\subseteq\hh_{b'}$ for all $0\leq b\leq b'\leq 1$. }
\label{fig:ellipses}
\end{figure}

Finally, we note that $\hh_b$ has a nesting property with respect to $b$: for all $0\leq b\leq b'\leq1$,
$$
B_{1/2}(0)\subseteq \hh_{b}\subseteq \hh_{b'}\subseteq B_{1}(0).
$$
We have already established the first and last inclusions. To see the middle inclusion, let
$\mvec(\xvec)=\svec(\nvec\cdot\xvec)$ be a negative energy spiral phase for $E_{b,\Hvecs}$. Let $\dvec=(n_1,bn_2)$ be the DMI vector associated with parameter $b$ and direction $\nvec$. Then there exists $\nvec'\in S^1$ such that $\dvec'=(n_1',b'n_2')=\alpha\dvec$ with $\alpha\geq 1$, see \figref{fig:ellipses}, so the DMI vector   associated with parameter $b'$ and direction $\nvec'$ is parallel to and longer than $\dvec$. But then the field $\mvec'(\xvec)=\svec(\nvec'\cdot\xvec)$ (with the same $\svec$, but oriented along $\nvec'$) is a spiral phase for $E_{b',\Hvecs}$ with the same exchange and potential energies as $\mvec$ but with DMI energy $E_{DMI,b'}(\mvec')=\alpha E_{DMI,b}(\mvec)\leq E_{DMI,b}(\mvec)$
since $E_{DMI,b}(\mvec)<0$. Hence $\mvec'$ is a negative energy spiral phase for 
$E_{b',\Hvecs}$.

For a fixed  $b\in[0,1]$ and field direction
$\hat\Hvec\in S^2$, we define the critical field
\beq
H_{SP}(b,\hat\Hvec)=\sup\{
|\Hvec|: |\Hvec|\hat\Hvec\in\hh_b
\}.
\eeq
This is the radius of $\hh_b$ along the ray containing $\hat\Hvec$.
As observed above, $1/2\leq H_{SP}\leq 1$. We now proceed to exhibit better lower bounds on $H_{SP}$ by exhibiting periodic fields of negative energy.

\subsection{The sine-Gordon bound}\label{sec:pendulum}

Configurations where the magnetization vector carries out a full rotation from the ferromagnetic vacuum $\hat\Hvec$ about an axis $\Nvec$ orthogonal to $\Hvec$ back to $\hat\Hvec$ feature prominently in the stability analysis of chiral magnet configurations \cite{Meynell2014_PRB, Muller2016_NJP, Kuchkin2020_PRB}, where they are called stripes. 
There are exact solutions of this nature in a variant of the model with a specifically tuned anisotropy potential 
\cite{barton2020magnetic}, which were used as a comparison configuration in  stability analysis of magnetic skyrmions in \cite{Ibrahim2023_CMP}. 

We are interested in periodic arrays of stripes. We begin 
by considering  a  spiral phase of the form 
\beq
\mvec(\xvec)=\svec(\nvec\cdot\xvec),\qquad
\svec(t) = \hat{\Hvec} \cos \theta(t) - \hat{\Hvec}\times {\Nvec} \sin \theta(t),
\label{eq:pendulum}
\eeq
where  $\Nvec\in\R^3$ is a unit vector orthogonal to $\hat\Hvec$, and $\theta$ a function to be determined. Extrema of $E$ restricted to this class of functions satisfy the static sine-Gordon equation
\beq
\ddot\theta=|\Hvec|\sin\theta.
\eeq
We will show that, for small enough $|\Hvec|$ such sine-Gordon fields exist with negative total energy, leading to a lower bound on $H_{SP}$ which we call the sine-Gordon bound
$H_*^{\rm pen}$.

For a fixed $\hat{\Hvec}\in S^2$ it is useful to define the quantity,
\beq
\delta(\hat{\Hvec}) := \max \left\{ 
\dvec(\nvec)\cdot \Nvec : 
\nvec \in S^1, 
\Nvec\in S^2,  
\Nvec\cdot\hat\Hvec=0
\right\}.
\label{eq:delta}
\eeq
This is the maximum of a smooth function over a compact manifold (a two-torus, in fact) and so certainly exists. We claim that, for all $b\in[0,1]$ and $\hat\Hvec\in S^2$,
\beq
H_{\text{\SP}}(b,\hat\Hvec) \geq H_\star^{\rm sG}:=\pi^2 \delta(\hat{\Hvec})^2/16.
\eeq

We establish this as follows. 
Given any fixed $\Hvec$, let $(\nvec, \Nvec) \in S^1 \times S^2$ be the point at which $\Nvec \cdot \hat{\Hvec} = 0$ and $\dvec(\nvec) \cdot \Nvec = \delta(\hat{\Hvec}).$
Consider the field \eqref{eq:pendulum}, 
where $\theta:\R\rightarrow \mathbb{R}$ is a smooth period $T$ function with $\theta(0) = 0$ and $\theta(T) = 2\pi$. This has average energy,
\begin{align}
\ip\ee & = \frac1T\int^T_0 \left\{ \frac{1}{2}\dot{\theta}^2 - \delta\dot{\theta} + |\Hvec|(1 - \cos\theta) \right\}\, dt\nonumber \\
& = \frac1T\left[- 2\pi \delta + \int^T_0 \left\{ \frac{1}{2}\dot{\theta}^2 + |\Hvec|(1-\cos\theta) \right\}\, dt\right].
\end{align}
We note that the static energy functional of the sine-Gordon model for a field in one dimension is,
\beq
E_{SG} = \int^\infty_{-\infty}\left\{\frac{1}{2} \dot{\theta}^2 + |\Hvec|(1-\cos\theta) \right\}\, dt,
\label{eq:sGE}
\eeq
which has the static kink solution \cite{manton2004topological},
\beq
\theta_K(t) = 4\tan^{-1}e^{t/\sqrt{|\Hvec|}},
\eeq
with the boundary conditions $\theta_K(-\infty) = 0$ and $\theta_K(\infty) = 2\pi$. Substituting this into \eqref{eq:sGE} gives total energy $E_{SG}(\theta_K) = 8\sqrt{|\Hvec|}$. Then for all $\eps > 0$, there exists $T$ sufficiently large and a map $\theta_\star:[0,T]\rightarrow \R$ with $\theta_\star(0) = 0$ and $\theta_\star(T) = 2\pi$ such that
\beq
\int^T_0\left\{ \frac{1}{2}\dot{\theta}^2_\star + |\Hvec|(1-\cos\theta_\star)\right\}\, dt < 8 \sqrt{|\Hvec|} + \eps.
\eeq
We can construct $\theta_\star$ by restricting the kink $\theta_K(t-T/2)$ to the interval $[0,T]$ then enforcing the boundary values with smooth cut-offs. The energy excess over $8\sqrt{|\Hvec|}$ can then be made arbitrarily small (by making $T$ sufficiently large), because $\theta_K$ has exponentially small tails.

Hence, for all $\eps > 0$, there exists a periodic field of type \eqref{eq:pendulum} with energy per period,
\beq
T\ip{E} < -2\pi \delta + 8\sqrt{|\Hvec|} + \eps.
\eeq
Choosing $\eps$ sufficiently small, this has $\ip\ee<0$ provided,
\beq
\sqrt{|\Hvec|} < \frac{\pi\delta}{4}.
\eeq
Hence, if $|\Hvec|<(\pi\delta(\hat\Hvec)/4)^2$, the system admits a negative energy 
spiral phase.

It would be useful to have an explicit formula for $H_\star^{\rm sG}(b,\hat\Hvec)$, but we have been unable to find one. One readily sees that
\beq
H_\star^{\rm sG}(b,(\pm 1,0,0))=\frac{\pi^2b^2}{16},\quad
H_\star^{\rm sG}(b,(0,\pm 1,0))=
H_\star^{\rm sG}(b,(0,0,\pm 1))=\frac{\pi^2}{16},
\eeq
and that 
\beq
H_\star^{\rm sG}(0,\hat\Hvec)=\frac{\pi^2}{16}(1-\hat H_1^2),\quad
H_\star^{\rm sG}(1,\hat\Hvec)=\frac{\pi^2}{16}.
\eeq
For $b$ small and $\hat\Hvec$ close to $\pm\evec_1$, the sine-Gordon bound is worse than the simple bound $H_{SP}\geq 1/2$. In this case, we can obtain a better lower bound on $H_{SP}$ by considering fields which precess in a spiral around $\Hvec$, as we now show.

\subsection{Conical bound}\label{sec:conical}
A further useful bound on the spiral domain can be obtained from trial configurations where the magnetization vector spirals on a cone. To define this, we choose  a unit vector $\uvec$ orthogonal to $\hat{\Hvec}$, an angular frequency $\omega \in \mathbb{R}$, angle $\theta \in [0,\pi]$ and unit vector $\nvec$ in the $x_1x_2$ plane  to construct the field,
\beq
\mvec(\xvec) = \hat{\Hvec}\cos\theta + (\uvec \cos(\omega \nvec \cdot \xvec) + \hat{\Hvec} \times \uvec \sin (\omega \nvec\cdot\xvec)) \sin\theta.
\label{spiralconf}
\eeq
As $t=\nvec\cdot \xvec$ varies, this maps out  a conical spiral with half apex angle $\theta$. Note that, for small $\theta$, such configurations can also be interpreted as approximate linearised configurations near the ferromagnetic vacuum $\nvec_0=\hat{\Hvec}$, since $\uvec$ and $\hat{\Hvec}\times \uvec$ span the tangent space to that vacuum state. The configurations are therefore closely related to those studied in \cite{barton2023stability}, where it was shown that linearised fields obey a gauged Klein-Gordon equation, with abelian gauge potential  $(a_1,a_2)=(\dvec_1\cdot \hat{\Hvec},\dvec_2\cdot \hat{\Hvec})$ in the notation of the current paper. The gauged Klein-Gordon equation acquires an imaginary mass when $a_1^2+a_2^2$ exceeds a bound proportional to the strength $H$ of the magnetic field, which was interpreted as an instability threshold for the ferromagnetic vacuum in \cite{barton2023stability}. In our notation, and restricted to the one dimensional reduction in the $\nvec$-direction, that bound can be re-interpreted as the condition
\beq
|\Hvec| < (a_1n_1+ a_2n_2)^2= (\dvec(\nvec)\cdot \hat{\Hvec)^2}
\label{concon}
\eeq
for conical spiral configurations like \eqref{spiralconf} to acquire negative energy per unit length.  We now derive this condition via a more direct route.

The average energy density of the field \eqref{spiralconf} is
\beq
\ip{\ee}=\frac12\sin^2\theta\omega^2+\dvec(\nvec)\cdot\hat\Hvec\omega\sin^2\theta-
+|\Hvec|(1-\cos\theta)
\eeq
which is minimized by choosing $\omega=-\dvec(\nvec)\cdot\hat\Hvec$, yielding
\beq
\ip{\ee}=-\frac12(\dvec(\nvec)\cdot\hat\Hvec)^2\sin^2\theta+|\Hvec|(1-\cos\theta).
\eeq
Minimizing this over $\theta\in[0,\pi]$, we see that
\beq
\ip{\ee}_{min}(\nvec)=\left\{\begin{array}{cl}
0, & \mbox{if $|\Hvec|\geq (\dvec(\nvec)\cdot\hat\Hvec)^2$,}\\
-\frac12(\dvec(\nvec)\cdot\hat\Hvec-|\Hvec|/\dvec(\nvec)\cdot\hat\Hvec)^2, &
\mbox{if $|\Hvec|\leq (\dvec(\nvec)\cdot\hat\Hvec)^2$,}
\end{array}\right.
\eeq
So there exists a negative energy spiral phase directed along $\nvec$ provided the condition \eqref{concon} holds.

The mapping $S^1\ra\R$,
\beq
\nvec\mapsto \dvec(\nvec)\cdot\hat\Hvec
\eeq
is maximized when $\nvec$ is directed along $(H_1,bH_2)$. Hence
\beq
H_{SP}(b,\hat\Hvec)\geq H_*^{\rm con}=\hat{H}_1^2+b^2\hat{H}_2^2,
\label{qhm}
\eeq
which coincides with \eqref{concon}.


Note that, for $\hat\Hvec=(\pm1,0,0)$ we have $H_{SP}\geq  H_*^{\rm con}=1$ from \eqref{qhm} and $H_{SP}\leq 1$ from \eqref{quhama}, so $H_{SP}=1$ in this case. So the intersection of $\hh_b$ with the $H_1$ axis is precisely the open interval $(-1,1)$.

\subsection{Finding the spiral phase numerically}

Recall that for any fixed $\nvec\in S^1$ and $T>0$, there is a spiral phase solution, translation invariant orthogonal to $\nvec$, periodic with period $T$ along $\nvec$, which minimizes $\ip\ee$ among all such fields. 
Its energy $\ip\ee$ will depend on $(\nvec,T)$. 
We define {\em the} spiral phase of the system to be the spiral phase solution which minimizes $\ip\ee$ among all possible choices of $(\nvec,T)$, if this exists. 

To find the spiral phase numerically, it is useful to rescale our ``time" coordinate so that $t=\nvec\cdot\xvec/T$, and $\svec(t)$ is now periodic with period $1$. The average energy density of a field $\mvec(\xvec)=\svec(\nvec\cdot\xvec/T)$ is
\beq
\ip{\ee}=\frac{E}{T}=\int_0^1\left\{
\frac{1}{2T^2}|\dot\svec|^2+\frac1T\dvec(\nvec)\cdot(\svec\times\dot\svec)+|\Hvec|-\Hvec\cdot\svec\right\}dt.
\eeq
We seek to minimize this over all unit period fields $\svec:\R/\Z\ra\ S^2$ and all directions $\nvec\in S^1$ and periods $T>0$. We will do this by iterating a two-step minimization process.

First we choose some $\nvec$, $T$ and $\svec:\R/\Z\ra S^2$, then minimize $\ip{\ee}(\svec)$ with respect to $\svec$, for $(\nvec,T)$ fixed, by arrested Newton flow. We will describe this computational method in more detail in \figref{sec:skylatt}. We then minimize $\ip{\ee}(\nvec,T)$ with respect to $(\nvec,T)$, for $\svec:\R/\Z\ra S^2$ fixed, by an explicit computation which we will describe shortly. We then iterate, alternating minimization steps with respect to the field and the direction/period, until $\ip{\ee}$ is minimized to a chosen tolerance (meaning the reduction in $\ip{\ee}$ in the last step is sufficiently small). 

So, for a fixed unit period field $\svec:\R/\Z\ra\R$, we must minimize $\ip{\ee}$ over $(\nvec,T)\in S^1\times(0,\infty)$. First note that
\beq
\ip{\ee}(\nvec,T)=\frac{K}{2T^2}+\frac1T\dvec(\nvec)\cdot\Svec +C,
\label{eq:theet}
\eeq
where
\bea
K&=&\int_0^1|\dot\svec(t)|^2\, dt, \nonumber \\
\Svec&=&\int_0^1\svec(t)\times\dot\svec(t)\, dt,\nonumber \\
C&=&\int_0^T V(\svec(t))\, dt
\eea
depend only on $\svec$. Only the second term in \eqref{eq:theet} depends on $\nvec$, and attains its minimum at
\beq\label{quehaymarn}
\nvec=-\frac{(S_1,b^2 S_2)}{\sqrt{S_1^2+b^2S_2^2}}.
\eeq
At this optimal $\nvec$,
\beq
\ip{\ee}(T)=\frac{K}{2T^2}-\frac1T\sqrt{S_1^2+b^2S_2^2}+C
\eeq
which attains its minimum at
\beq\label{quehaymarT}
T=\frac{K}{\sqrt{S_1^2+b^2S_2^2}}.
\eeq
Hence, the direction $\nvec$ and period $T$ which minimize $\ip{\ee}$ for a given unit period field $\svec$ are \eqref{quehaymarn}, \eqref{quehaymarT} respectively. 

This scheme allows us to construct, for each $\hat\Hvec$ and $b$, the energetically optimal spiral phase, and to map out the parameter space on which this phase has negative average energy, and hence is favoured over the FM state. In this way, we can construct $\hh_b$ numerically.

\begin{figure}
\begin{center}
\includegraphics[width=\linewidth,trim={0 4cm 0 0},clip]{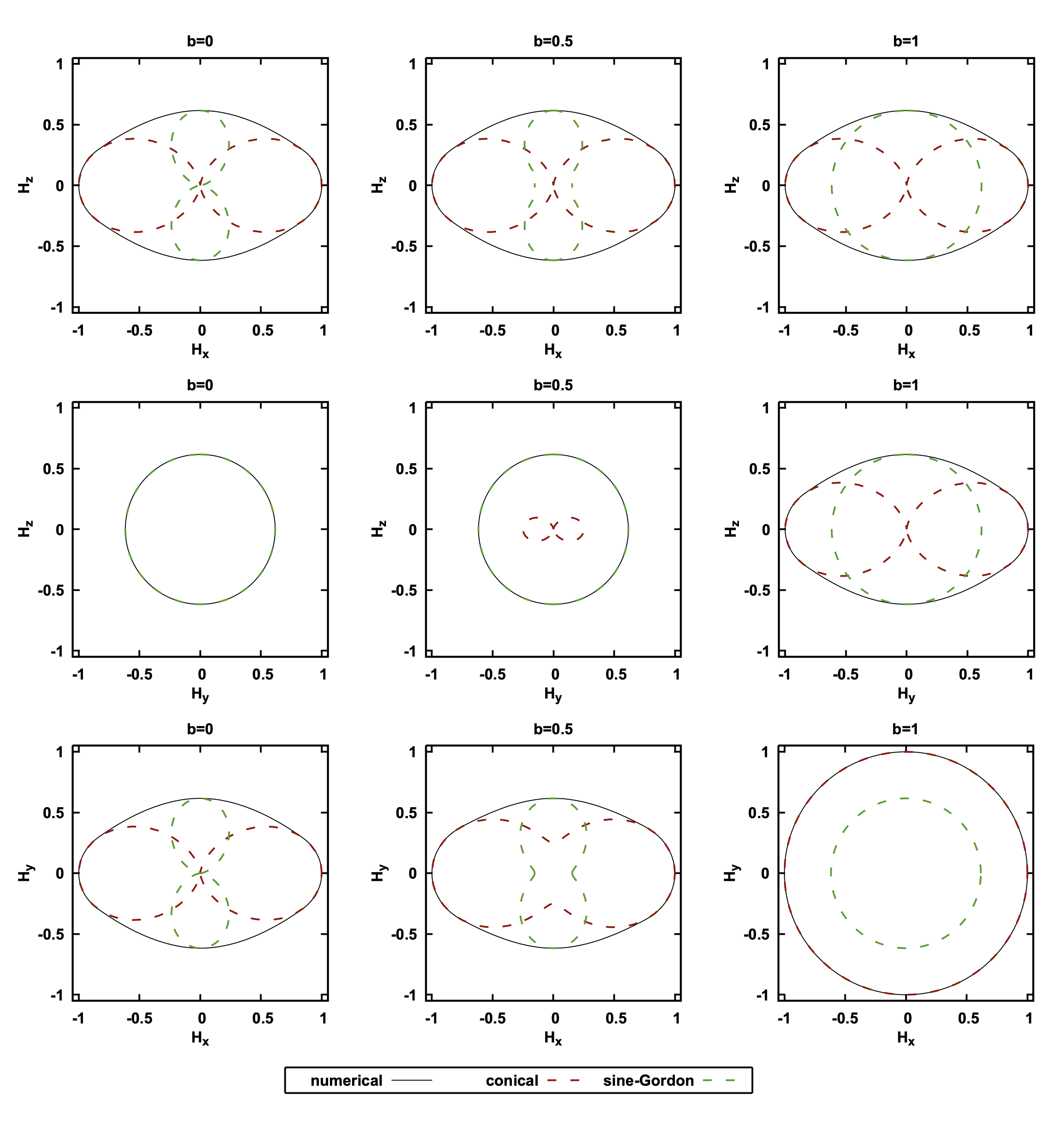}
\end{center}
\caption{Sections through the spiral domain $\hh_b$ in the coordinate planes
$H_2=0$ (top row), $H_1=0$ (middle row) and $H_3=0$ (bottom row) for DMI parameters $b=0$ (the rank 1 case, left column), $b=0.5$ (middle column) and $b=1$ (the standard isotropic case, right column). In each case, $\hh_b$ is the region bounded by the black closed curve and is, by definition, the set of applied field values
$\Hvec$ for which the model has a negative energy spiral phase.  The dashed curves correspond to lower bounds on the radius of the boundary curve, calculated by assuming the SP state is a conical spiral (red) or sine-Gordon state (green).}
\label{fig:SPTransitions}
\end{figure}

The \figref{fig:SPTransitions} shows sections through $\hh_b$ along the coordinate planes for $b=0,0.5,1$, generated using this numerical scheme. In each case
$\hh_b$ is the region interior to the black closed boundary curve. The dashed curves are lower bounds on the radius of this boundary curve obtained by considering conical spiral phases (red) and sine-Gordon phases (green), as explained in sections \ref{sec:conical} and \ref{sec:pendulum}.
Note that these numerical results confirm all the properties of $\hh_b$ derived above:
$\hh_0$ and $\hh_1$ are volumes of revolution symmetric under rotations about the $H_1$ and $H_3$ axes respectively, all sections are star-shaped and inversion symmetric, and for all sections $\hh_0\subseteq \hh_{0.5}\subseteq\hh_1$, consistent with the nesting property. We observe that for $\Hvec$ in the $H_1H_3$ plane, the energetically optimal spiral phase is always aligned with the $x_1$ axis, that is, has $\nvec=(1,0)$. The DMI vector $\dvec((1,0))=(1,0,0)$ for all $b\in[0,1]$, so we find that the $H_2=0$ section of $\hh_b$ (the top row of \figref{fig:SPTransitions}) is independent of $b$. While this property is plausible on symmetry grounds, we have been unable to derive it rigorously.

Particular examples of optimal spiral phases may be seen in later figures, combined with skyrmion lattices.

\section{Finding skyrmion lattices}\label{FSL}\label{sec:skylatt}\news

Skyrmion lattices are solutions $\mvec:\R^2\ra S^2$ of the Euler-Lagrange equations for the energy functional $E$ \eqref{eq:E} which are doubly periodic with respect to some linearly independent pair of vectors $\vvec_1,\vvec_2\in\R^2$, that is,
\beq
\mvec(\xvec+n_1\vvec_1+n_2\vvec_2)=\mvec(\xvec)
\eeq
for all $(n_1,n_2)\in\Z^2$ and all $\xvec\in\R^2$. We may equivalently regard $\mvec$ as a solution on the flat torus $\Omega=\R^2/\Lambda$ where 
\beq
\Lambda=\{n_1\vvec_1+n_2\vvec_2:(n_1,n_2)\in\Z^2\}<\R^2
\eeq
is the lattice generated by $\{\vvec_1,\vvec_2\}$.


Recalling that our field $\mvec$  represents the magnetization of a very large but bounded thin film ferromagnet, we now seek lattice solutions which minimise the total energy of our large but finite sample. Since the total energy is, trivially, the energy per unit area multiplied by the fixed total area, this amounts to searching for pairs $(\mvec,\Lambda)$ of  lattice configurations  and period lattices which minimize {\em energy per unit area} among all nearby combinations of field configuration and period lattice

To do this in practice, we start with  a fixed lattice $\Lambda$, and  an initial field $\mvec_0:\Omega=\ra S^2$ of nonzero degree $Q$,  and flow by energy gradient descent to a local energy minimizer among fields in its homotopy class. It is an interesting and, so far as we are aware, open problem to find necessary and sufficient conditions on $\Lambda$ and $Q$ under which such a minimizer exists. However, for small $|Q|$ we expect energy minimizers to exist for almost all choices of $\Lambda$. Most of these solutions have no physical significance: they are artifacts of our choice of boundary conditions that will never be observed in a real system. To represent a genuine physical skyrmion lattice, our map
$\mvec:\Omega \ra S^2$ should minimize energy per unit area  not only with respect to variations of the field $\mvec$ on the fixed torus $\Omega$, but also with respect to variations of the period lattice $\Lambda$.

For a given a lattice $\Lambda$ generated by $\vvec_1,\vvec_2$, it is convenient to use coordinates $X_1,X_2\in[0,1]$ defined so that
\beq
\xvec=X_1\vvec_1+X_2\vvec_2.
\eeq
In matrix language, $\xvec=L\Xvec$ where $L\in GL(2,\R)$ is the invertible matrix with columns $\vvec_1,\vvec_2$, and $\Xvec=M\xvec$, where $M=L^{-1}$. Note that the area of the torus $\Omega=\R^2/\Lambda$ is $|\det L|$.
The energy per unit area of a field $\mvec:\Omega\ra S^2$ is then
\beq\label{mw}
\ip{\ee}(M)=\frac{1}{2}\tr(M^THM)+\tr(\DD M)+C
\eeq 
where 
\bea
H_{ij}&=&\int_{\Box}\frac{\cd m_a}{\cd X_i}\frac{\cd m_a}{\cd X_j}dX_1dX_2 \label{hdef}\\
\DD_{ij}&=&D_{ai}\epsilon_{abc}\int_{\Box}m_b\frac{\cd m_c}{\cd X_j}dX_1dX_2
\label{ddef}\\
C&=&\int_\Box V(\mvec)dX_1dX_2
\eea
and $\Box=[0,1]\times[0,1]$ is the closed unit square. The expression for $\ip\ee$ in \eqref{mw} is simply \eqref{eq:E} integrated over $\Omega$ with a change of basis and divided by the area of $\Omega$, namely $|\det L|$.

Shifting perspective slightly, we may interpret \eqref{mw} as follows: all two-tori $\R^2/\Lambda$ are diffeomorphic through linear diffeomorphisms, so we may identify them all with the single reference torus $\Omega_\Box=\R^2/\Z^2$ and regard all doubly periodic fields as maps $\mvec:\Omega_\Box\ra S^2$. If we do so, our energy functional $E$ now depends parametrically on the lattice $\Lambda$ or, equivalently, on the invertible matrix
$M$, as in \eqref{mw}. For a given field $\mvec:\Omega_\Box\ra S^2$, we seek the minimum of $\ip{\ee}:GL(2,\R)\ra\R$. We could do this numerically by a gradient descent method (for example, arrested Newton flow) on $GL(2,\R)$. This is the approach taken by Lee {\it et al}.\  in \cite{leetosmunkwo} to construct skyrmion lattices in a specific planar model with axisymmetric DMI term and a novel potential.  In fact, it is straightforward to compute the minimizer of
$\ip{\ee}:GL(2,\R)\ra\R$ explicitly, as we now observe, and we find this to be more computationally efficient.

The minimum must occur at a critical point of $\ip{\ee}$. Now the differential of $\ip{\ee}$ at $M\in GL(2,\C)$ is the linear map
\beq
d\ip{\ee}_M:\eps\mapsto\tr(\eps^T(HM+\DD^T))
\eeq
and $M$ is a  critical point of $\ip{\ee}$ if and only if this vanishes on all $2\times 2$ matrices $\eps$, that is, if
\beq
HM+\DD^T=0.
\eeq
The matrix $H$ is manifestly symmetric and positive semi-definite. If the map $\mvec$ is once continuously differentiable and has any regular points (points $\Xvec\in \Omega_\Box$ at which $d\mvec_{\vecs{X}}$ has maximal rank), it is positive definite, and hence invertible. Every map of degree $Q\neq 0$ has regular points, so if $Q\neq 0$, $H$ is invertible. Then $\ip{\ee}$ has a unique critical point
\beq
M_*=-H^{-1}\DD^T,
\eeq
which must be a global minimum since $\ip{\ee}$ at large 
$|M|$ is dominated by the positive quadratic term $\frac12\tr(M^THM)$. Note that
$\det M_*=\det\DD/\det H$, so strictly speaking, $\ip{\ee}$ attains a minimum if and only if $\DD$ is invertible. 

We claim that $\DD$ is generically invertible (that is, is invertible for almost all maps $\mvec$) if $D$ has rank 2, but is generically
{\em not} invertible if $D$ has rank 1. To see this, consider the linear map
\beq
Z:\R^2\ra\R^3,\qquad Z(V)=\int_\Box\mvec\times \left(V_1\frac{\cd \mvec}{\cd X_1}
+V_2\frac{\cd \mvec}{\cd X_2}\right)dX_1dX_2,
\eeq
noting that, for generic $\mvec$, this has rank 2. Now assume that $\DD$ is not invertible. Then there exists $V\in\R^2$, $V\neq 0$, such that $\DD V=0$. But
\bea
(\DD V)_i
=-\dvec_i\cdot Z(V).
\eea
Hence, $\det\DD=0$ if and only if there exists $V\in\R^2$ such that $Z(V)$ is perpendicular to the span of $\{\dvec_1,\dvec_2\}$. If $D$ has rank 2, this is equivalent to the plane $Z(\R^2)$ containing the vector $\dvec_1\times\dvec_2$ which is generically false. Hence $\DD$ is generically invertible. If $D$ has rank 1, however, the span of $\{\dvec_1,\dvec_2\}$ is one dimensional, so its orthogonal complement intersects every two-plane in $\R^3$. Hence $\DD$ is generically not invertible in this case. This reinforces our expectation, motivated by the anisotropic Derrick scaling argument presented in section \ref{rank1}, that rank 1 materials do not support skyrmion lattices. 

In summary, the energetically optimal lattice for a given field $\mvec:\R^2/\Z^2\ra S^2$ has period matrix
\beq
L=-\DD^{-1}H,
\eeq
assuming, as is generically true, that both $H$ and $\DD$ are invertible. This observation provides the basis for our numerical method for finding skyrmion lattices. We start by choosing an initial smooth field $\mvec_0:\Omega_\Box\ra S^2$ of degree $Q<0$, small (typically we consider only $Q=-1$ and $Q=-2$), and a matrix $M_0\in GL(2,\R)$ close to $\I_2$, chosen at random. We then evolve $\mvec(t)$ by arrested Newton flow for $E(\mvec,M_0)$, with $M$ fixed, until $\mvec(t)$ converges to $\mvec_1$ say, a local minimum of $E(\cdot, M_0)$. (The arrested Newton flow algorithm will be described in detail shortly.) We then construct the matrices
$H(\mvec_1)$, $\DD(\mvec_1)$, as in \eqref{hdef}, \eqref{ddef}, and compute
$M_1=-H^{-1}\DD^T$. We now iterate this two step process: we perform arrested Newton flow starting at $\mvec_n$ for the energy functional $E(\cdot,M_n)$, converging to a local minimum $\mvec_{n+1}$, then construct 
$M_{n+1}=-H(\mvec_{n+1})^{-1}\DD(\mvec_{n+1})^T$. In this way we construct a sequence of pairs $(\mvec_n,M_n)$ with monotonically decreasing $\ip{\ee}$, terminating the process when $\ip{\ee}_{n-1}-\ip{\ee}_n$ falls below some tolerance.
Note that, during this process we may generate fields for which $\det\DD(\mvec_n)<0$, in which case $\det M_n<0$ and the associated linear diffeomorphism
$M_n^{-1}:\Omega_\Box\ra\Omega_n$ is orientation {\em reversing}. In this case, the topological degree of $\mvec_n$, regarded (as it physically is) as a map from
$\Omega_n$ (the torus with period matrix $L_n=M_n^{-1}$) to $S^2$ is actually $-Q$, not $Q$. So our scheme will automatically flip the sign of $Q$ if, for the given spiralization tensor $D$, that is energetically favourable. 

To minimize $\ip\ee$ with respect to $\mvec:\R^2/\Z^2\ra S^2$ we use arrested Newton flow \cite{spewin}. First we discretize the unit square so that $(X_1,X_2)$ takes values on a $N\times N$ lattice of spacing $h=1/N$, with periodic boundary conditions. Our magnetization field is now approximated by $N^2$ points $\mvec_{i,j}\in S^2$. By approximating spatial derivatives by finite differences (we use standard 4th order approximations) we obtain a discrete approximant of our energy functional $\ip\ee_{dis}: (S^2)^{N^2}\ra \R$. We seek minima of this function. To find them, we solve for the motion of a notional point particle in
$\CC_{dis}=(S^2)^{N^2}$ moving under the potential energy $\ip\ee_{dis}$ according to Newton's law 
\beq
\ddot{\mvec} = -\grad \ip{\ee}_{dis}\left( \mvec \right),
\eeq
starting at some initial configuration $\mvec(0)$ with $\dot{\mvec}(0) = \zerovec$. This time evolution is computed approximately using the standard 4th order Runge-Kutta scheme with fixed time step $\delta t=h/2$. As time evolves, the  ``particle" runs downhill in the energy landscape defined by $\ip\ee_{dis}$, but since Newton flow conserves total energy (meaning $\ip\ee_{dis}+\frac12|\dot\mvec|^2$) it will not, without intervention, converge to a minimum of $\ip\ee_{dis}$. This intervention is the arresting criterion: at each time step, we check whether $\dot\mvec(t+\delta t)\cdot\grad\ip\ee_{dis}(\mvec(t+\delta t))>0$. If so, the ``particle" has started to travel uphill, so we arrest the flow, that is, set $\dot\mvec(t+\delta t)=\zerovec$ and restart the flow at $\mvec(t+\delta t)$ \cite{speight2021magnetic}. The scheme halts when $\mvec$ is a critical point of $\ip\ee_{dis}$ to within some tolerance $\eps$, meaning that
$\|\grad\ip\ee_{dis}(\mvec)\|_\infty<\eps$, where $\|\cdot\|_{\infty}$ denotes the sup norm on $\R^{3N^2}$.

The above second order flow converges to a critical point consistently much faster than evolving the physically realistic dynamics governed by the first order damped LLG equation. Clearly, the path constructed $\mvec(t)$ has no dynamical significance, but as a tool to obtain static solutions, it is significantly more efficient than LLG flow.

\begin{figure}
\begin{center}
\includegraphics[width=0.9\linewidth]{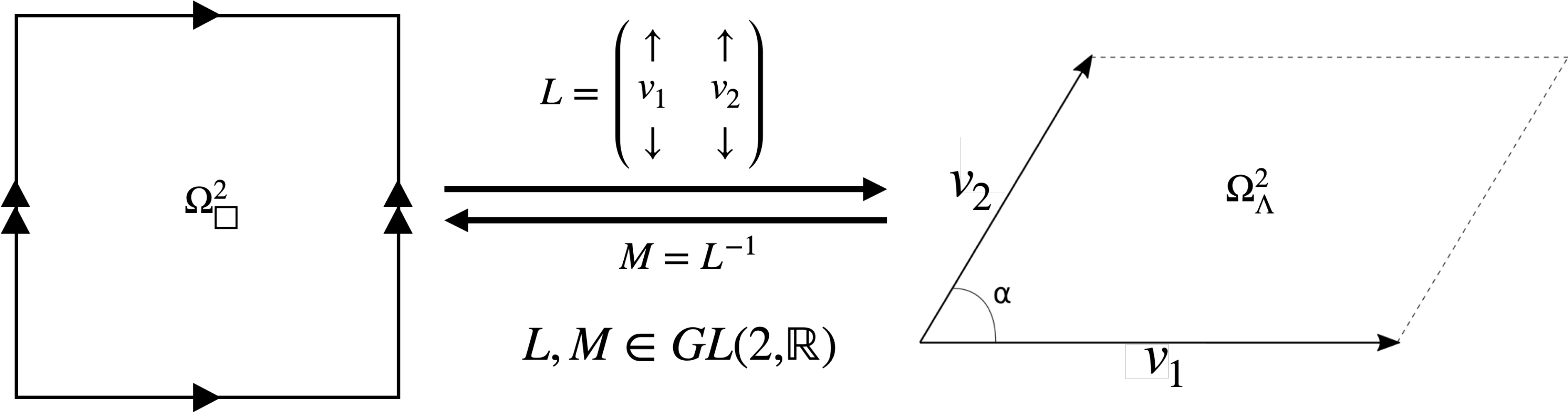}
\end{center}
\caption{\label{fig:unitCell}
Diagram of the geometry of a general unit cell $T^2_\Omega$ for a skyrmion lattice, defined by the two vectors $\vvec_1$ and $\vvec_2$ with angle $\alpha$. Note that $\vvec_1$, $\vvec_2$ need not be oriented and if $\det M < 0$ then $\mvec(x)$ has degree $N = - N_\square$.}
\end{figure}

\section{Results}

We now consider the phase space for chiral lattices in two dimensions. We will break it into three sections: the ferromagnetic phase (FM), the spiral phase (SP) and skyrmion lattice phase (SK).  To find the phase for a particular value of parameters we attempt to find global minimizers for one dimensional and two dimensional periodic systems using the methods outlined in the previous two sections.  We then compare the energy per unit area of these solutions with the zero energy of the FM state and the lowest of the three gives the phase. 

The full parameter space is four dimensional, consisting of the applied field  $\Hvec \in \mathbb{R}^3$ and the DMI parameter $b \in [0,1]$.  Systematically exploring this space requires a lot of simulations.


\begin{figure}
\begin{center}
\begin{overpic}[width=0.5\linewidth,trim={0.53cm 20.6 0 0},clip]{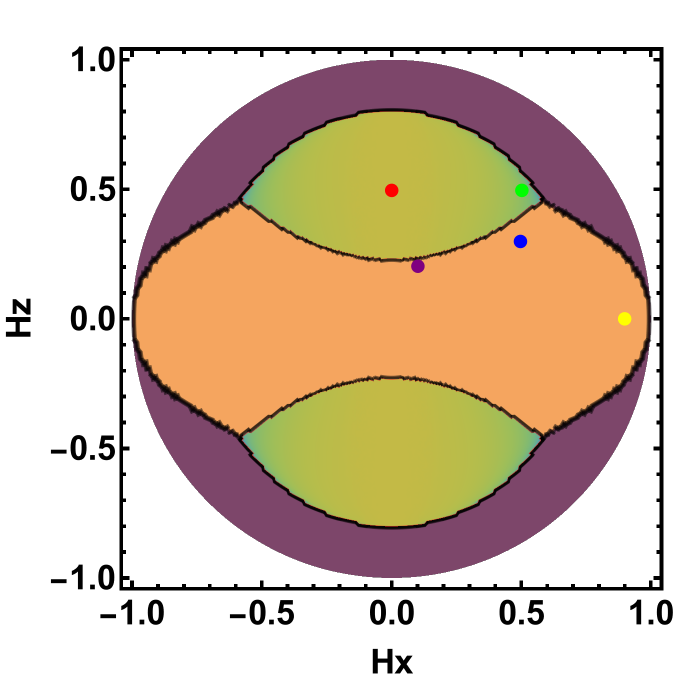}
\put(51,-5){\Large $\boldsymbol{H_\perp}$}
\put(-5,48){\Large $\boldsymbol{H_3}$}
\put(49.5,71){$\boldsymbol{A}$}
\put(70,71){$\boldsymbol{B}$}
\put(70,57){$\boldsymbol{C}$}
\put(85,46){$\boldsymbol{D}$}
\put(53.5,54){$\boldsymbol{E}$}
\end{overpic}\includegraphics[width=0.095\linewidth,trim={0 -15 0 0},clip]{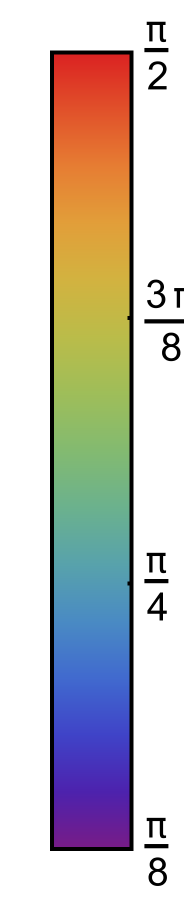}
\end{center}
\caption{Phase diagram for the model with $b=1$ (the standard DMI term) and varying applied field $\Hvec$. Since this model has rotational symmtery about the $H_3$ axis, we may represent it using $H_3$ and $H_\perp$, the component of $\Hvec$ orthogonal to $(0,0,1)$. The diagram is coloured based to represent the solution with lowest energy per unit area. The phases are ferromagnetic (FM) - purple; Spiral phase (SP) - orange; and skyrmion lattice (SK) - coloured by the interior angle of the unit cell with colours given by the colour bar. The energetically optimal solutions at the labelled points are shown in \figref{fig:fixedDMIconfigs}.}
\label{fig:fixDMI}
\end{figure}

\begin{figure}
\begin{center}
\begin{overpic}[width=0.3\linewidth]{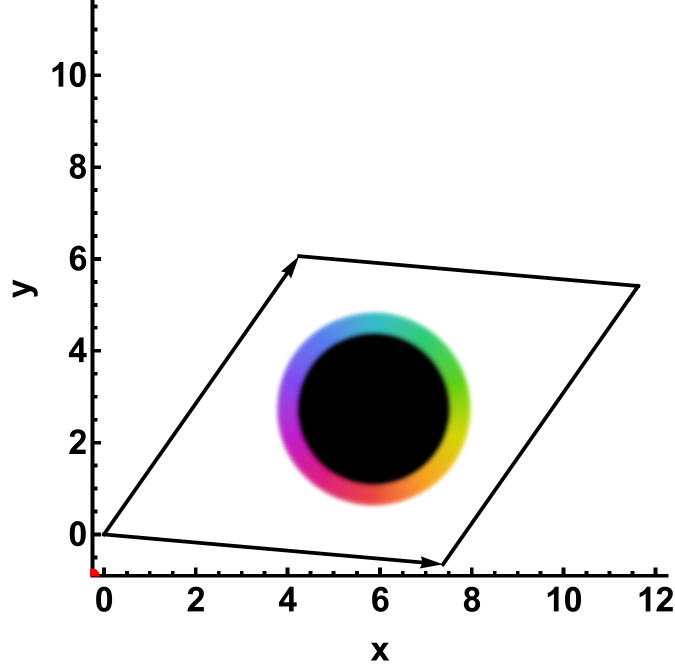}
\put(-1,95){$\boldsymbol{A}$}
\put(11,12){\tikzcircle[red]{3pt}}
\end{overpic}
\begin{overpic}[width=0.3\linewidth]{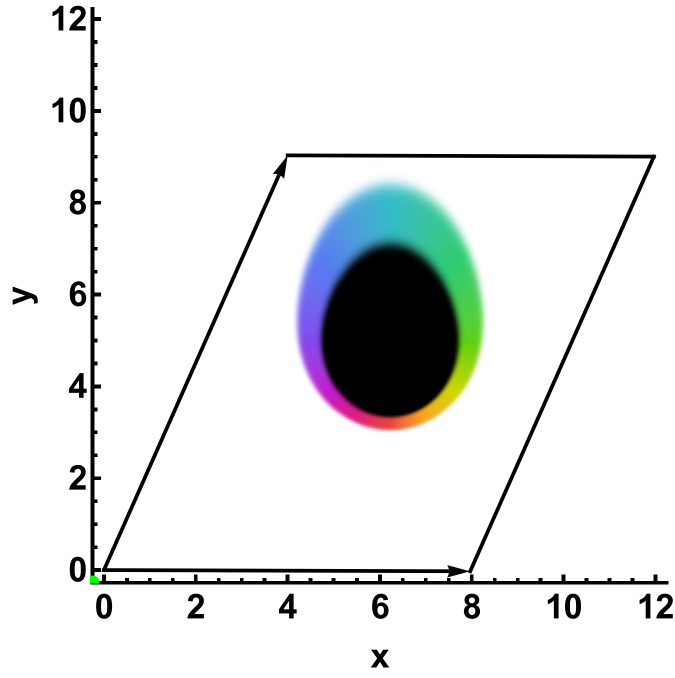}
\put(-1,95){$\boldsymbol{B}$}
\put(11,11.3){\tikzcircle[green]{3pt}}
\end{overpic}
\begin{overpic}[width=0.3\linewidth]{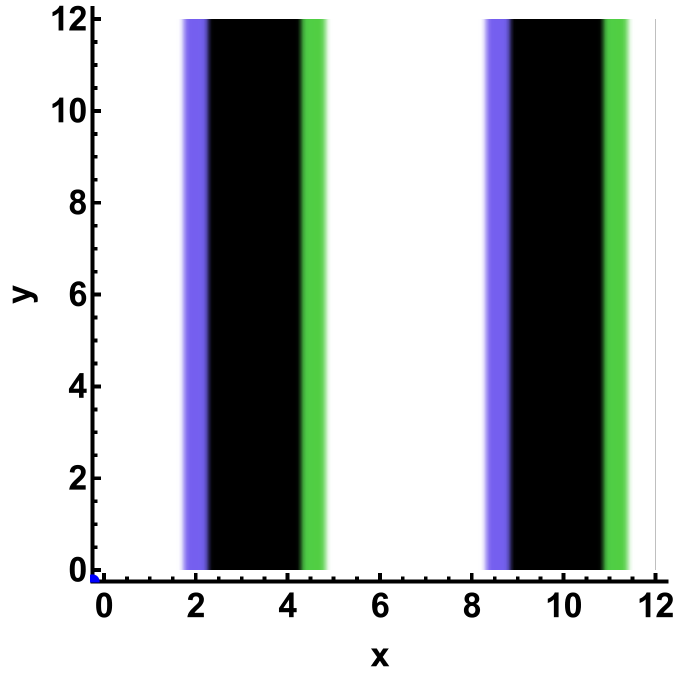}
\put(-1,95){$\boldsymbol{C}$}
\put(11,11.3){\tikzcircle[blue]{3pt}}
\end{overpic}
\begin{overpic}[width=0.3\linewidth]{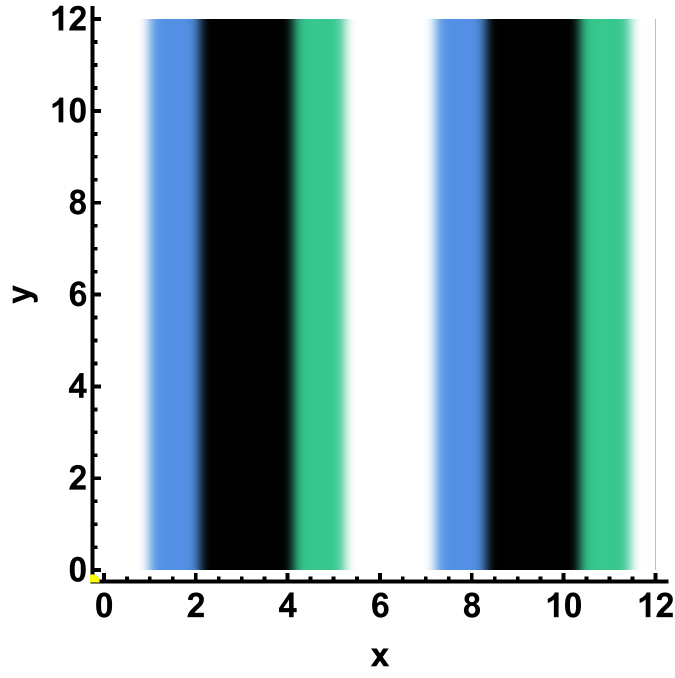}
\put(-1,95){$\boldsymbol{D}$}
\put(11,11.3){\tikzcircle[yellow]{3pt}}
\end{overpic}
\begin{overpic}[width=0.3\linewidth]{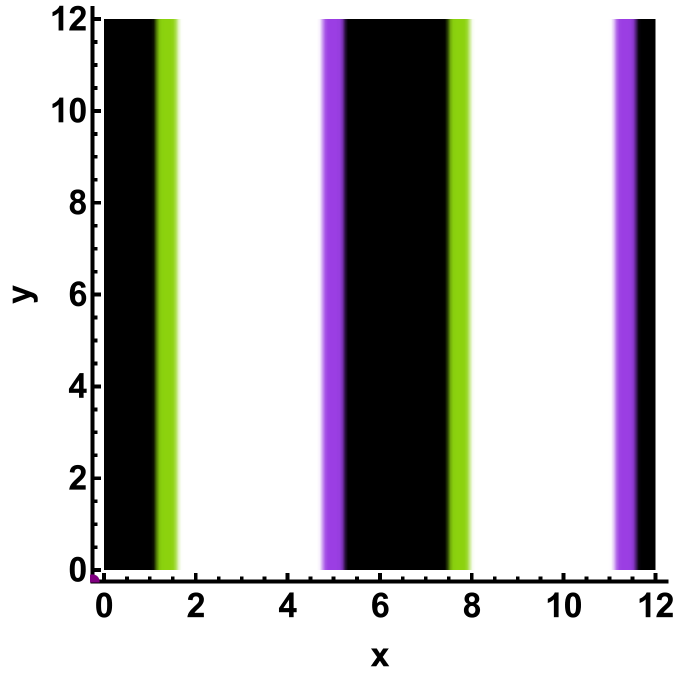}
\put(-1,95){$\boldsymbol{E}$}
\put(11,11.3){\tikzcircle[violet]{3pt}}
\end{overpic}
\end{center}
\caption{Five optimal configurations for DMI term $b=1$ and applied field in the $x-z$ plane, marked on the phase diagram in \figref{fig:fixDMI}. For skyrmion lattices (A,B) the unit cell's period vectors are shown and spiral states (C,D,E) have translation invariance. 
The colouring convention used to represetn $\mvec$ is explained 
 in \figref{fig:skyrmion}.}
\label{fig:fixedDMIconfigs}
\end{figure}

We will start by considering the standard DMI term $b=1$ and varying the applied field direction, the result of which can be seen in \figref{fig:fixDMI}.  Note that for $b=1$ the model has a mixed rotation symmetry about the $z$-axis, see equation \eqref{1sym}, so we may assume, without loss of generality that $\Hvec=(H_1,0,H_3)$. In the plot the purple region is the FM state, the orange is the SP state and the other region is the SK state, coloured based on the angle of the unit cell. It is unsurprising that in all directions for strong/weak applied field we get the FM/SP state, as the potential/gradient terms dominate. This is a feature of all the simulations we ran.  In addition, as we apply an increasingly stronger field in a given direction if the angle away from the $H_3$-axis is $\theta \gtrsim \pi/4$ then we have only a single phase transition from SP to FM, matching the discussion in the section \ref{SP}. 
This means that for $\theta \lesssim \pi/4$ we have two phase transitions from the SP state to the SK state and then from the SK state to the FM state.  We also see that as we move further from the $H_\perp = 0$ line, the unit cell of the SK state is increasingly distorted. The skyrmion lattices at field values marked $A,B$ and the spiral phases at values marked $C,D,E$ on figure \figref{fig:fixDMI} are presented in figure \figref{fig:fixedDMIconfigs}, using the colouring scheme depicted in figure \figref{fig:skyrmion}. Note that the spiral phase is always oriented along the $x_1$ axis.


\begin{figure}
\begin{center}
\begin{overpic}[width=0.48\linewidth,trim={20 20 0 -25},clip]{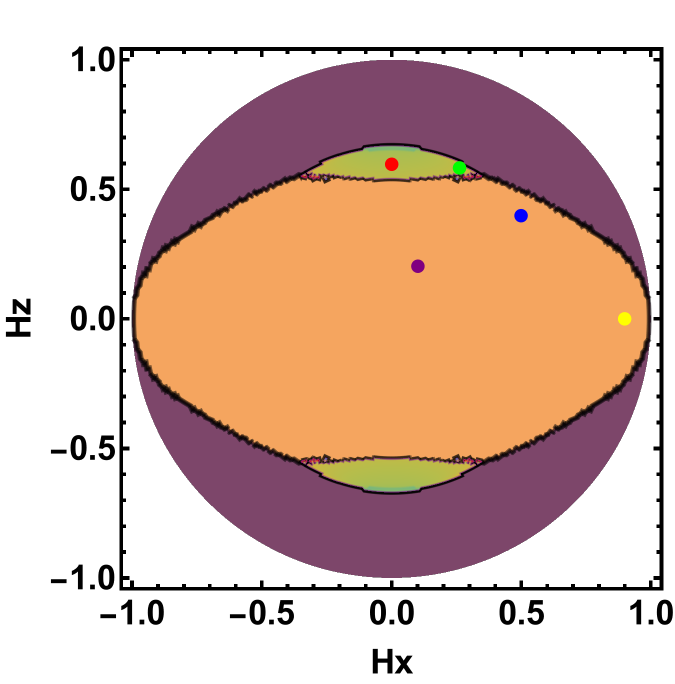}
\put(-6,44){\Large $\boldsymbol{H_3}$}
\put(41,89){\Large $\boldsymbol{\phi = 0^\circ}$}
\put(44.5,68){$\boldsymbol{A}$}
\put(56,63){$\boldsymbol{B}$}
\put(63,58){$\boldsymbol{C}$}
\put(78,42){$\boldsymbol{D}$}
\put(48,50){$\boldsymbol{E}$}
\end{overpic}
\begin{overpic}[width=0.48\linewidth,trim={20 20 0 -25},clip]{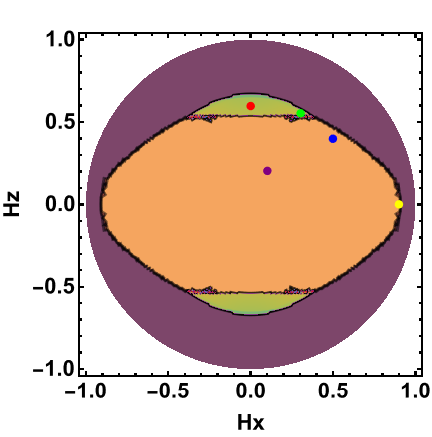}
\put(41,88){\Large $\boldsymbol{\phi = 30^\circ}$}
\put(45,68){$\boldsymbol{A}$}
\put(56,62.5){$\boldsymbol{B}$}
\put(63.5,57.5){$\boldsymbol{C}$}
\put(78.5,41.5){$\boldsymbol{D}$}
\put(49,49.5){$\boldsymbol{E}$}
\end{overpic}
\begin{overpic}[width=0.48\linewidth,trim={20 20 0 -25},clip]{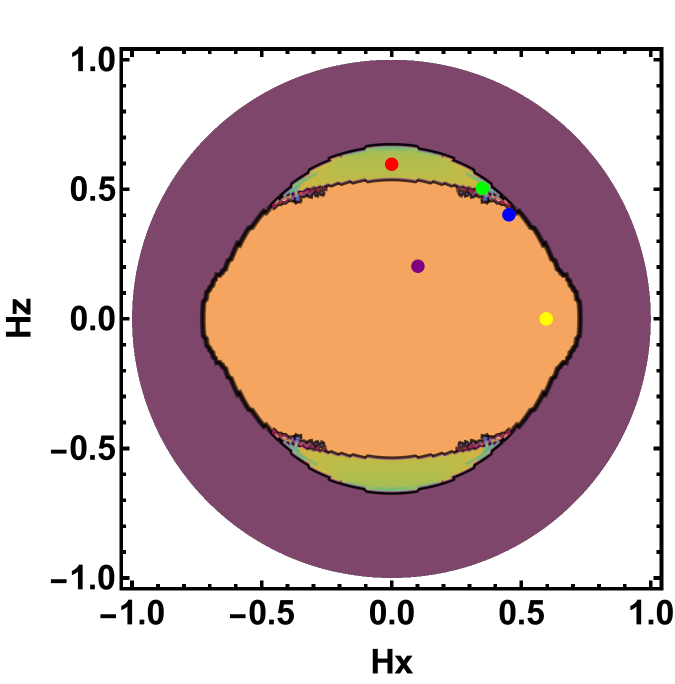}
\put(46,-5){\Large $\boldsymbol{H_\perp}$}
\put(-6,44){\Large $\boldsymbol{H_3}$}
\put(41,89){\Large $\boldsymbol{\phi = 60^\circ}$}
\put(44.5,67.6){$\boldsymbol{A}$}
\put(57,62){$\boldsymbol{B}$}
\put(62,57){$\boldsymbol{C}$}
\put(66,42){$\boldsymbol{D}$}
\put(48,50){$\boldsymbol{E}$}
\end{overpic}
\begin{overpic}[width=0.48\linewidth,trim={20 20 0 -25},clip]{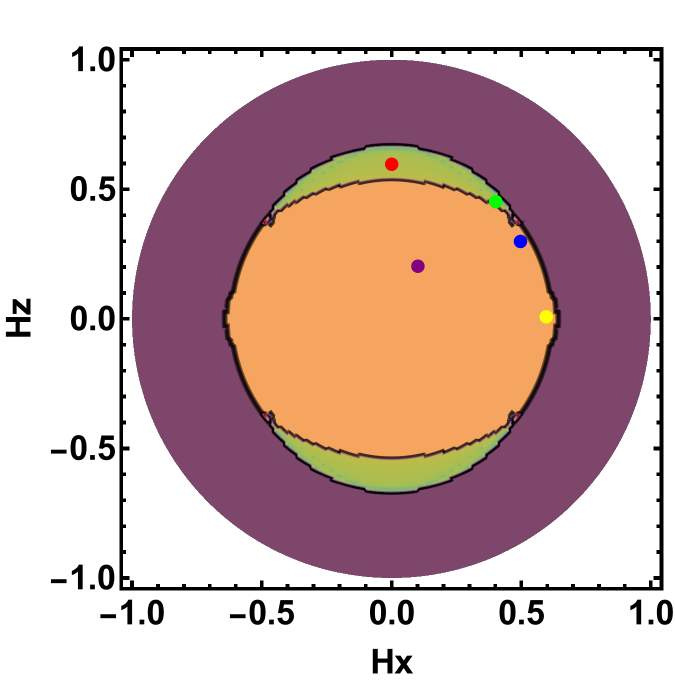}
\put(46,-5){\Large $\boldsymbol{H_\perp}$}
\put(41,89){\Large $\boldsymbol{\phi = 90^\circ}$}
\put(44.5,67.6){$\boldsymbol{A}$}
\put(58.5,60.5){$\boldsymbol{B}$}
\put(62.6,53.5){$\boldsymbol{C}$}
\put(66,42.5){$\boldsymbol{D}$}
\put(48,50){$\boldsymbol{E}$}
\end{overpic}
\end{center}
\caption{Phase diagrams for the model with DMI parameter $b=0.8$ and varying applied field $\Hvec = |H_3|\hat{e}_3 + H_\perp$ in different planes defined by $\hat{H}_\perp = (\cos \phi, \sin \phi, 0)$. As $b \neq 1$, symmetry about the $z$-axis is broken. The colouring conventions are the same as \figref{fig:fixDMI}.
The energetically optimal solutions at the labeled points are shown in \figref{fig:fixDMIb0p8}.}
\label{fig:fixDMI2}
\end{figure}

\begin{center}
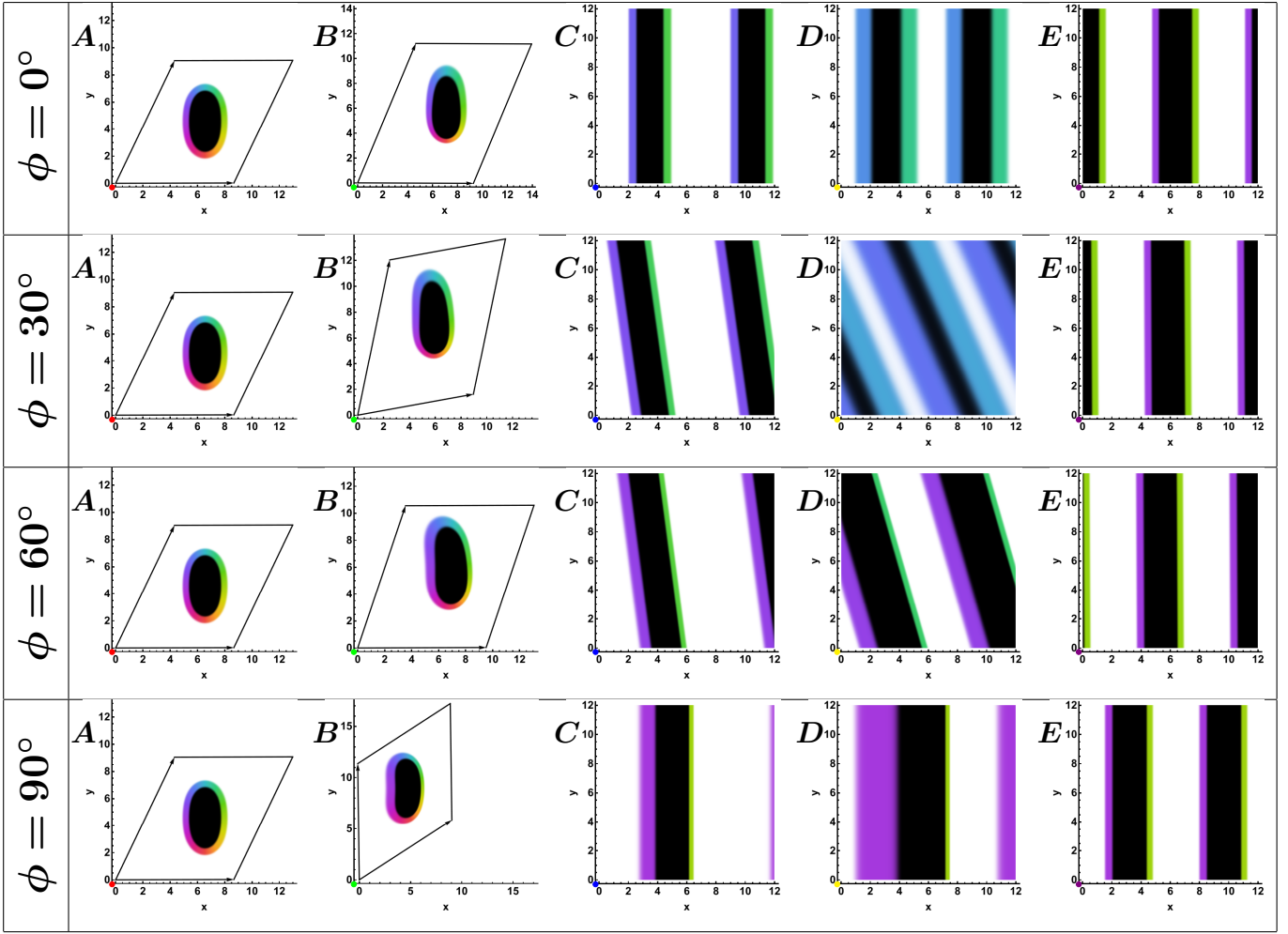
\begin{figure}
\makebox[\textwidth][c]{
\begin{tabular}{|c|ccccc|}
\hline
{\Large \parbox[c][.8cm][c]{.5cm}{\centering\rotatebox{90}{$\qquad\qquad\;\;\boldsymbol{\phi=0^\circ}$}}}
&\begin{overpic}[width=0.18\linewidth]{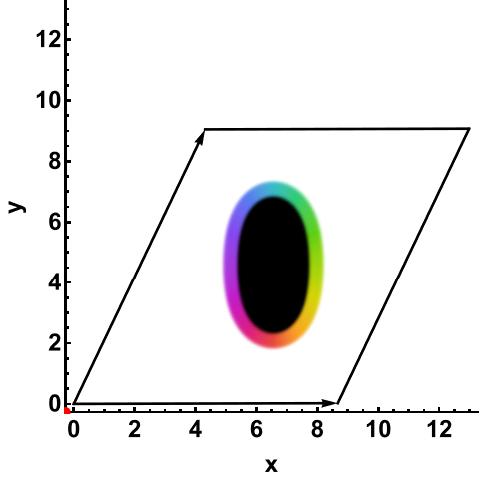}
\put(12,10.8){\tikzcircle[red]{1pt}}
\put(-6,80){$\boldsymbol{A}$}
\end{overpic}&
\begin{overpic}[width=0.18\linewidth]{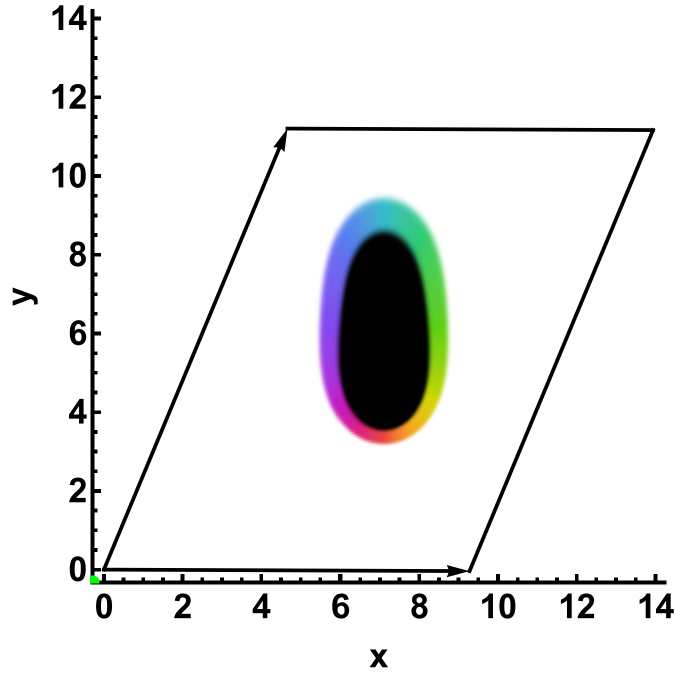}
\put(12,10.8){\tikzcircle[green]{1pt}}
\put(-6,80){$\boldsymbol{B}$}
\end{overpic}&
\begin{overpic}[width=0.18\linewidth]{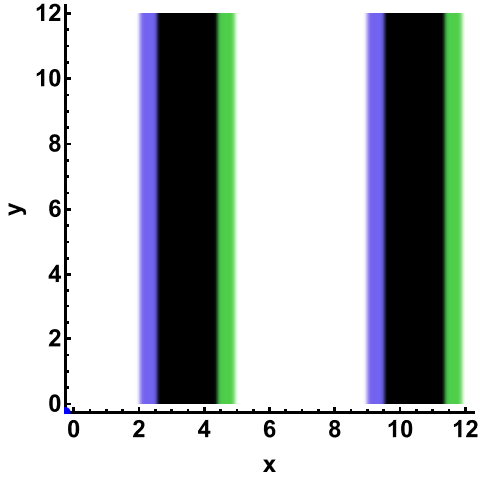}
\put(12,10.8){\tikzcircle[blue]{1pt}}
\put(-6,80){$\boldsymbol{C}$}
\end{overpic}&
\begin{overpic}[width=0.18\linewidth]{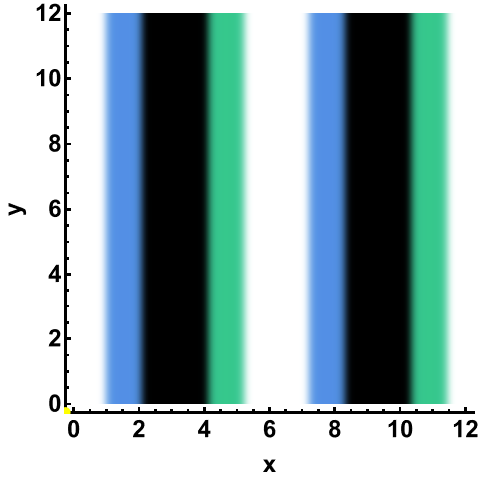}
\put(12,10.8){\tikzcircle[yellow]{1pt}}
\put(-6,80){$\boldsymbol{D}$}
\end{overpic}&
\begin{overpic}[width=0.18\linewidth]{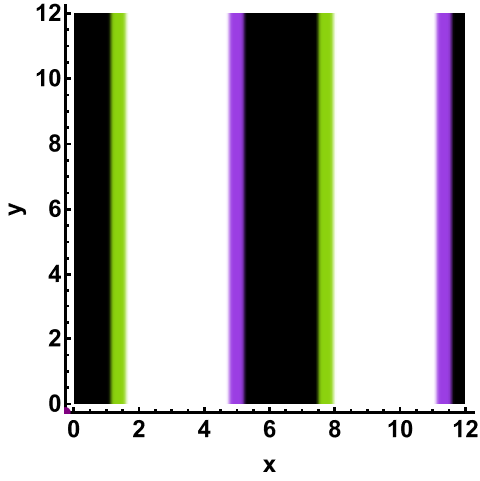}
\put(12,10.8){\tikzcircle[violet]{1pt}}
\put(-6,80){$\boldsymbol{E}$}
\end{overpic}\\
\hline
{\Large \parbox[c][.8cm][c]{.5cm}{\centering\rotatebox{90}{$\qquad\qquad\;\,\boldsymbol{\phi=30^\circ}$}}}&
\begin{overpic}[width=0.18\linewidth]{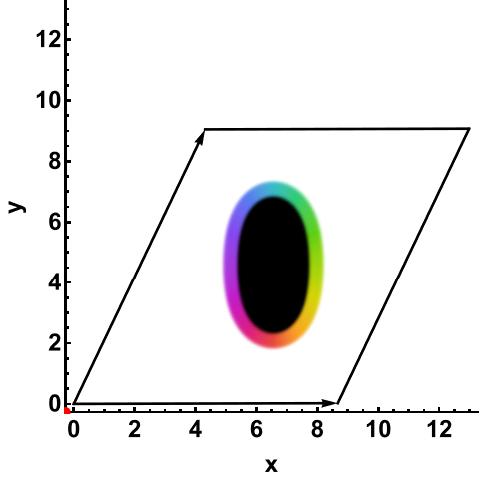}
\put(12,10.8){\tikzcircle[red]{1pt}}
\put(-6,80){$\boldsymbol{A}$}
\end{overpic}&
\begin{overpic}[width=0.18\linewidth]{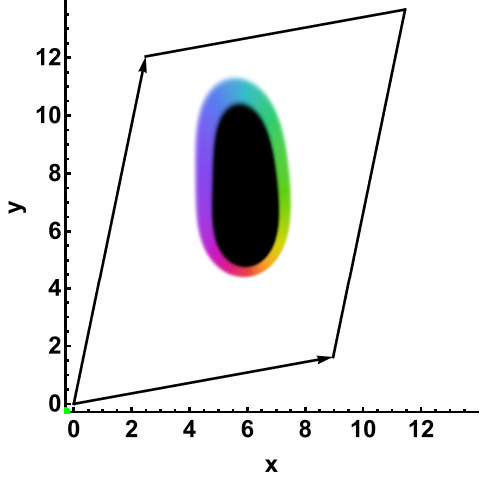}
\put(12,10.8){\tikzcircle[green]{1pt}}
\put(-6,80){$\boldsymbol{B}$}
\end{overpic}&
\begin{overpic}[width=0.18\linewidth]{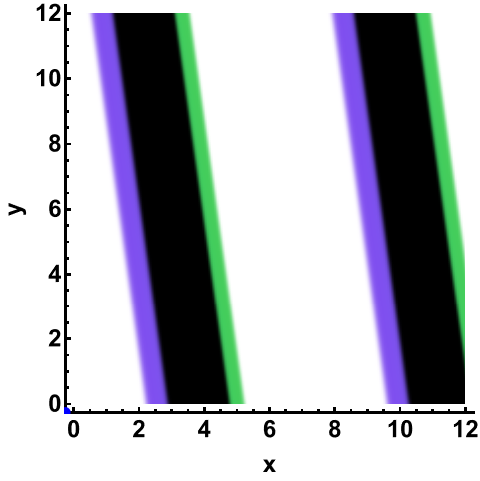}
\put(12,10.8){\tikzcircle[blue]{1pt}}
\put(-6,80){$\boldsymbol{C}$}
\end{overpic}&
\begin{overpic}[width=0.18\linewidth]{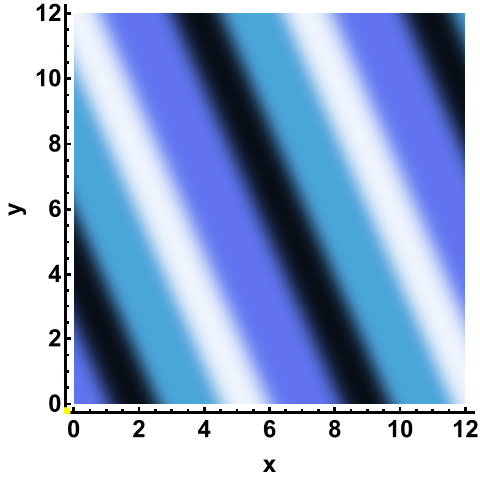}
\put(12,10.8){\tikzcircle[yellow]{1pt}}
\put(-6,80){$\boldsymbol{D}$}
\end{overpic}&
\begin{overpic}[width=0.18\linewidth]{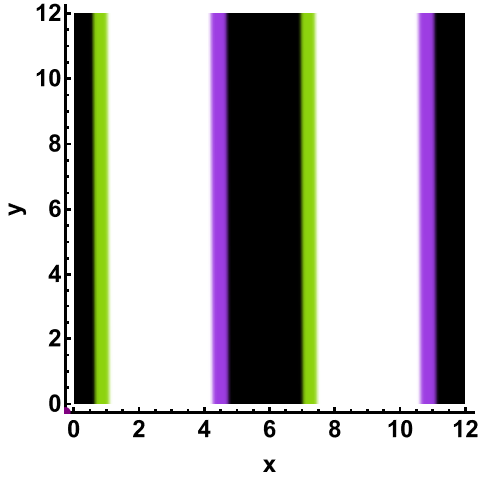}
\put(12,10.8){\tikzcircle[violet]{1pt}}
\put(-6,80){$\boldsymbol{E}$}
\end{overpic}\\
\hline
{\Large \parbox[c][.8cm][c]{.5cm}{\centering\rotatebox{90}{$\qquad\qquad\;\,\boldsymbol{\phi=60^\circ}$}}}&
\begin{overpic}[width=0.18\linewidth]{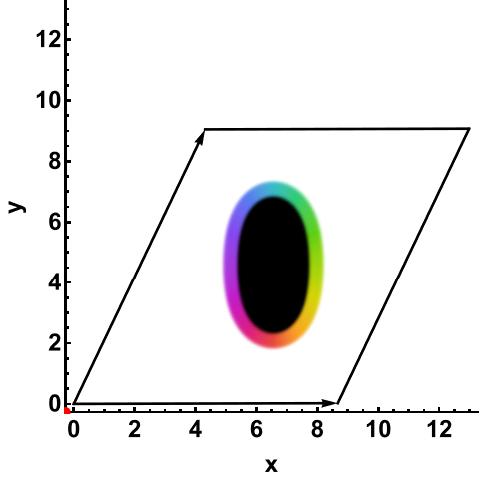}
\put(12,10.8){\tikzcircle[red]{1pt}}
\put(-6,80){$\boldsymbol{A}$}
\end{overpic}&
\begin{overpic}[width=0.18\linewidth]{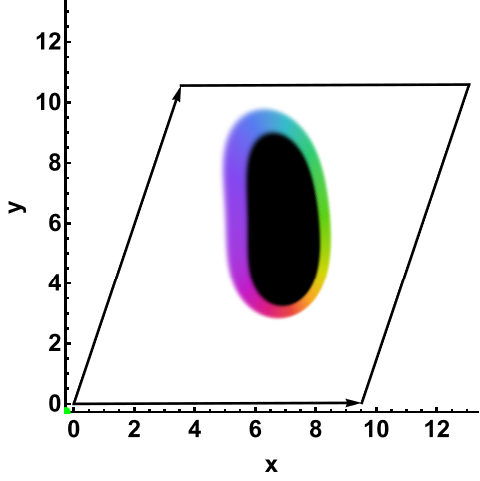}
\put(12,10.8){\tikzcircle[green]{1pt}}
\put(-6,80){$\boldsymbol{B}$}
\end{overpic}&
\begin{overpic}[width=0.18\linewidth]{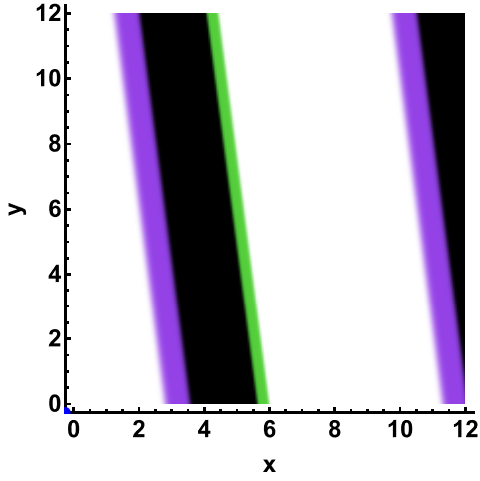}
\put(12,10.8){\tikzcircle[blue]{1pt}}
\put(-6,80){$\boldsymbol{C}$}
\end{overpic}&
\begin{overpic}[width=0.18\linewidth]{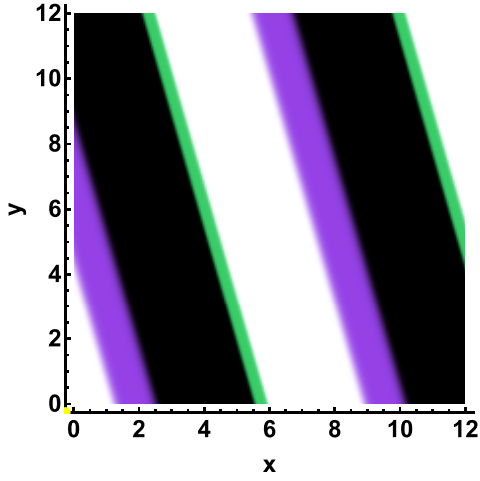}
\put(12,10.8){\tikzcircle[yellow]{1pt}}
\put(-6,80){$\boldsymbol{D}$}
\end{overpic}&
\begin{overpic}[width=0.18\linewidth]{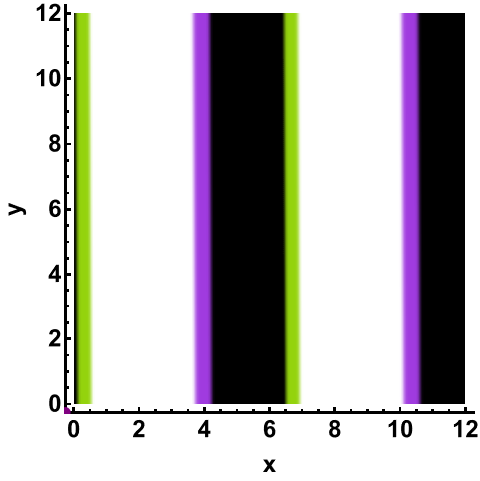}
\put(12,10.8){\tikzcircle[violet]{1pt}}
\put(-6,80){$\boldsymbol{E}$}
\end{overpic}\\
\hline
{\Large \parbox[c][.8cm][c]{.5cm}{\centering\rotatebox{90}{$\qquad\qquad\;\,\boldsymbol{\phi=90^\circ}$}}}&
\begin{overpic}[width=0.18\linewidth]{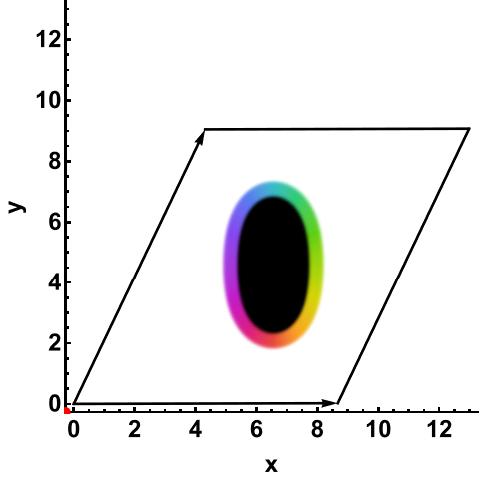}
\put(12,10.8){\tikzcircle[red]{1pt}}
\put(-6,80){$\boldsymbol{A}$}
\end{overpic}&
\begin{overpic}[width=0.18\linewidth]{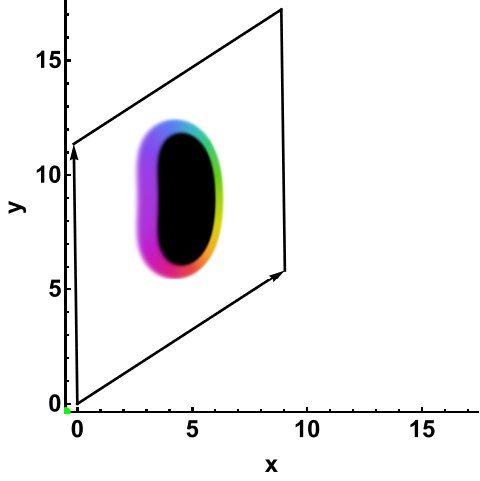}
\put(12,10.8){\tikzcircle[green]{1pt}}
\put(-6,80){$\boldsymbol{B}$}
\end{overpic}&
\begin{overpic}[width=0.18\linewidth]{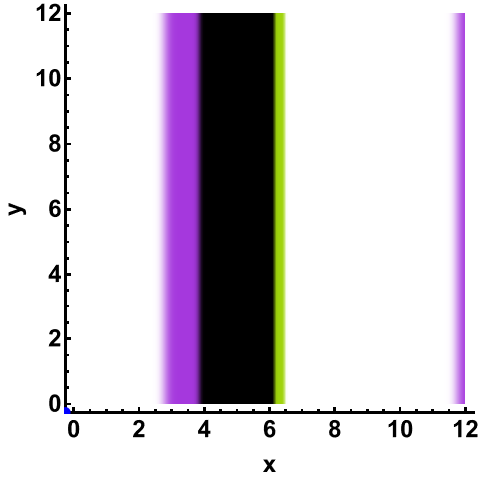}
\put(12,10.8){\tikzcircle[blue]{1pt}}
\put(-6,80){$\boldsymbol{C}$}
\end{overpic}&
\begin{overpic}[width=0.18\linewidth]{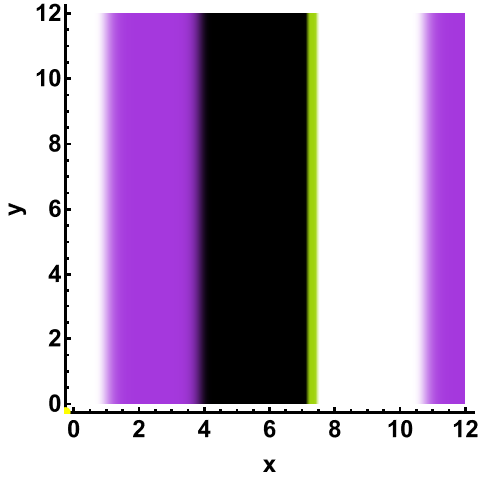}
\put(12,10.8){\tikzcircle[yellow]{1pt}}
\put(-6,80){$\boldsymbol{D}$}
\end{overpic}&
\begin{overpic}[width=0.18\linewidth]{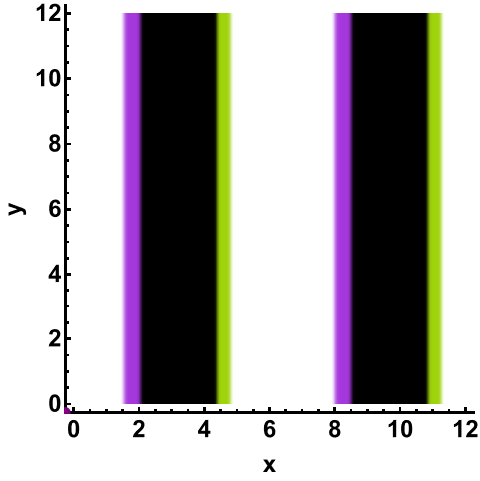}
\put(12,10.8){\tikzcircle[violet]{1pt}}
\put(-6,80){$\boldsymbol{E}$}
\end{overpic}\\
\hline
\end{tabular}
}
\caption{Five optimal configurations for DMI term $b=0.8$ with applied field in $H_3-H_\perp$ plane, where $H_\perp = (\cos \phi, \sin\phi,0)$ planes, marked on the corresponding phase diagram for $\phi$ in \figref{fig:fixDMI2}. For skyrmion lattices (A,B) the unit cell's period vectors are shown and spiral states (C,D,E) have translation invariance in differing directions. 
The colouring convention used to represent $\mvec$ is explained 
 in \figref{fig:skyrmion}.}
\label{fig:fixDMIb0p8}
\end{figure}
\end{center}

We next consider the model with $b=0.8$, which has no rotation symmetry. To fully appreciate the phase diagram, we consider a sequence of two dimensional sections sections through it, where $\Hvec=|\Hvec|(\sin\theta\cos\phi, \sin\theta\sin\phi, \cos\theta)$ with the azimuthal angle $\phi$ fixed. In particular, we consider sections with $\phi\in\{0, \pi/6, \pi/3, \pi/2\}$, see \figref{fig:fixDMIb0p8}. Note that the volume of the SK phase has shrunk quite considerably  
compared with the $b=1$ system, and that the SP domain shrinks as we increase $\phi$. Again, a selection of skyrmion lattices and spiral phases at marked points in figure \figref{fig:fixDMIb0p8} are displayed in figure \figref{fig:fixDMIb0p8}. We see that in this less symmetric model, there is nontrivial interplay between the orientation of $\Hvec$ and the optimal DMIN vector $\dvec(\nvec)$ for spiral phases, so that the spiral phases are not uniformly oriented along the $x_1$ axis.


\begin{figure}
\begin{center}
\begin{overpic}[width=0.5\linewidth,trim={0.53cm 20.6 0 0},clip]{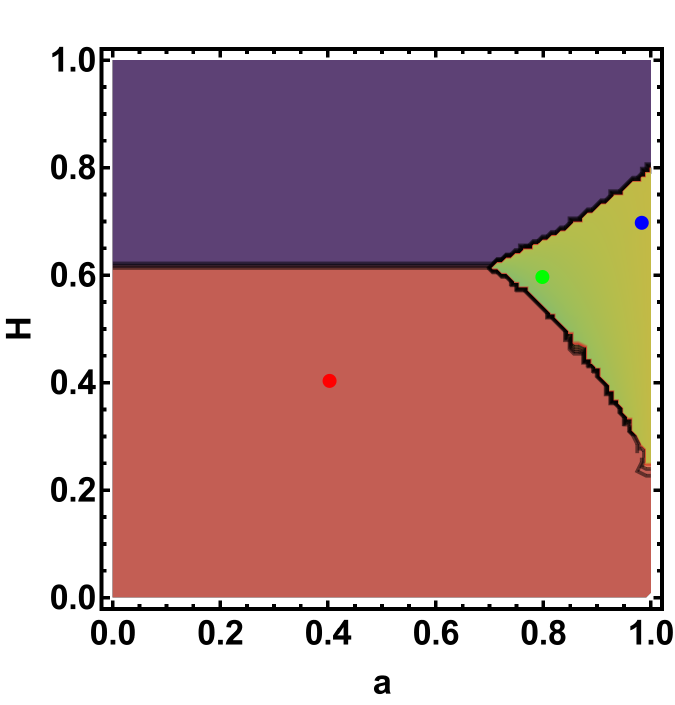}
\put(51,-4){\Large $\boldsymbol{b}$}
\put(-10,48){\Large $\boldsymbol{|H|}$}
\put(45,46){$\boldsymbol{A}$}
\put(78,54){$\boldsymbol{B}$}
\put(86,61){$\boldsymbol{C}$}
\end{overpic}\includegraphics[width=0.095\linewidth,trim={0 -15 0 0},clip]{colBar.pdf}
\end{center}
\caption{Phase diagram for the model with varying DMI parameter $b\in[0,1]$ and applied field strength $|\Hvec|$ for fields directed along $\hat\Hvec=(0,0,1)$. 
The colouring conventions are the same as in \figref{fig:fixDMI}. Note that $b=1$ corresponds to the standard DMI term, while $b=0$ corresponds to the rank $1$ case. Note also that the skymrion lattice phase vanishes completely for $b<b_*\approx 0.71$ . 
The energetically optimal solutions at the labeled points are shown in
\figref{fig:fixedHconfigs}.}
\label{fig:fixH}
\end{figure}

\begin{figure}
\begin{center}
\begin{overpic}[width=0.245\linewidth,trim={20 20 0 0},clip]{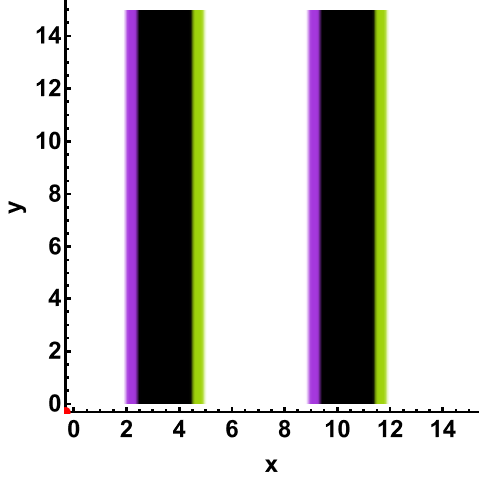}
\put(6.2,7.4){\tikzcircle[red]{3pt}}
\put(-8,83){$\boldsymbol{A}$}
\end{overpic}
\begin{overpic}[width=0.245\linewidth,trim={20 20 0 0},clip]{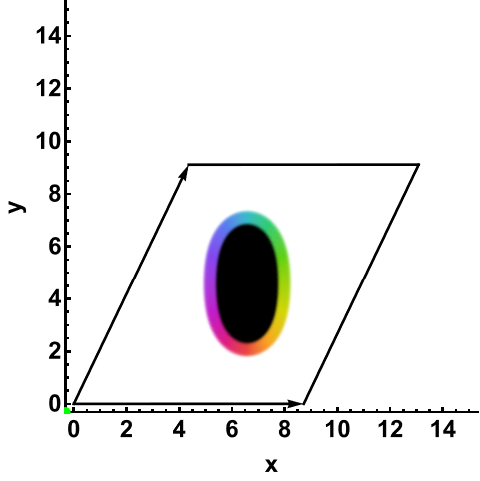}
\put(6.2,7.4){\tikzcircle[green]{3pt}}
\put(-8,83){$\boldsymbol{B}$}
\end{overpic}
\begin{overpic}[width=0.245\linewidth,trim={20 20 0 0},clip]{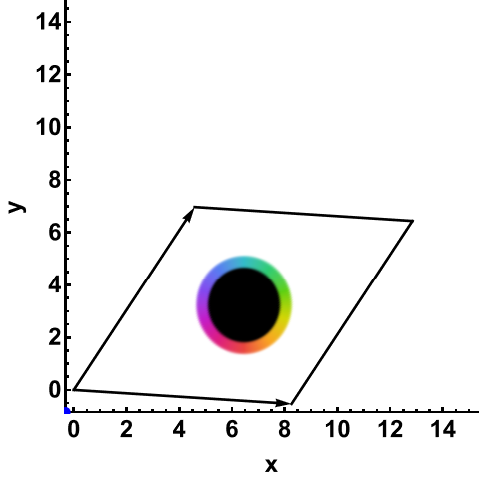}
\put(6,7.4){\tikzcircle[blue]{3pt}}
\put(-8,83){$\boldsymbol{C}$}
\end{overpic}
\end{center}
\caption{Three optimal configurations for applied field in the $\hat{z}$ direction with different DMI terms as marked on the phase diagram in \figref{fig:fixH}. For skyrmion lattices (B,C) the unit cell with period vectors is shown and the spiral state (A) has translation invariance. The colouring of the configurations is explained in \figref{fig:skyrmion}.}
\label{fig:fixedHconfigs}
\end{figure}
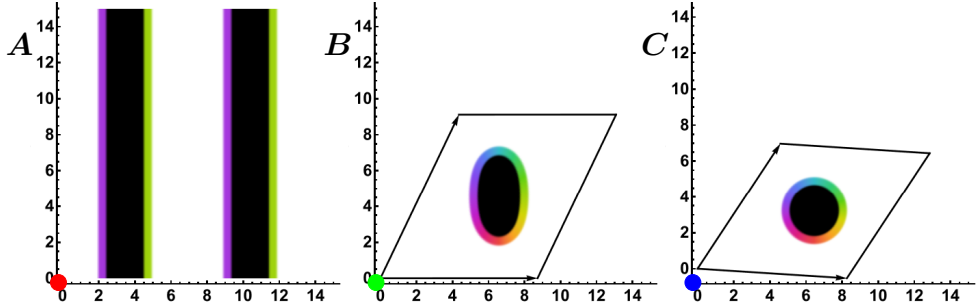

We have seen that the volume of the SK domain in $\Hvec$ space for $b=0.8$ is considerably smaller than that for $b=1$. The next results present phase diagrams in the $(b,|\Hvec|)$ plane for a collection of five fixed applied field directions $\hat\Hvec$, starting with $\hat\Hvec=(0,0,1)$, see figure \figref{fig:fixH}. Once again, the energetically optimal fields at the marked points are depicted in \figref{fig:fixedHconfigs}. Note that the SK domain vanishes completely for
$b$ sufficiently small. This is consistent with our earlier observations (sections \ref{rank1} and \ref{FSL}) that the rank 1 model (with $b=0$) should have no stable skyrmion lattices, but it is quite surprising that the SK phase vanishes at
$b\approx 0.71$, which is rather large. As we will see, this critical value of $b$ only increases if we choose $\hat\Hvec\neq(0,0,1)$. 
So skyrmion lattices exist as the ground state only for models with DMI parameter close to $1$, and then only for applied fields close to being orthogonal to the plane spanned by the DMI vectors (which is the $(m_1,m_2)$ plane in our choice of coordinates). Skyrmion lattices are thus surprisingly fragile structures.


\begin{figure}
\begin{tabular}{cc}
\begin{overpic}[width=0.48\linewidth,trim={20 20 0 0},clip]{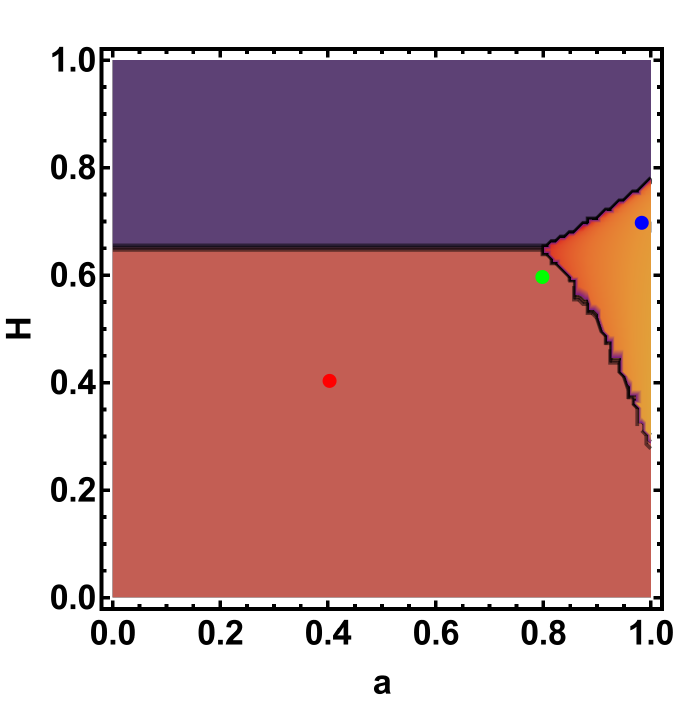}
\put(-13,49){\Large $\boldsymbol{|H|}$}
\put(40,98){\Large $\boldsymbol{\phi = 0^\circ}$}
\put(44.5,44.5){$\boldsymbol{A}$}
\put(69,55){$\boldsymbol{B}$}
\put(84,60.5){$\boldsymbol{C}$}
\end{overpic}&
\begin{overpic}[width=0.48\linewidth,trim={20 20 0 0},clip]{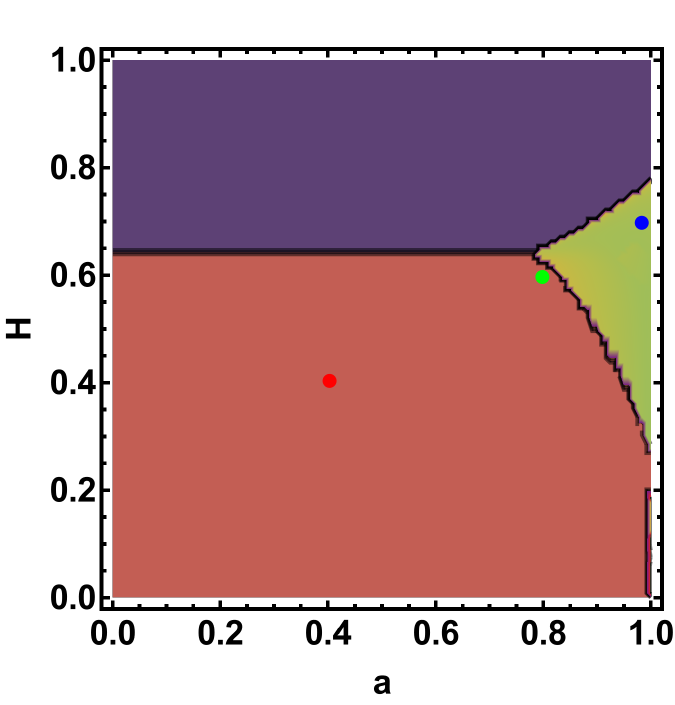}
\put(40,98){\Large $\boldsymbol{\phi = 30^\circ}$}
\put(44.5,44.5){$\boldsymbol{A}$}
\put(69,55){$\boldsymbol{B}$}
\put(84,60.5){$\boldsymbol{C}$}
\end{overpic}\vspace{.5cm}\\
\begin{overpic}[width=0.48\linewidth,trim={20 20 0 0},clip]{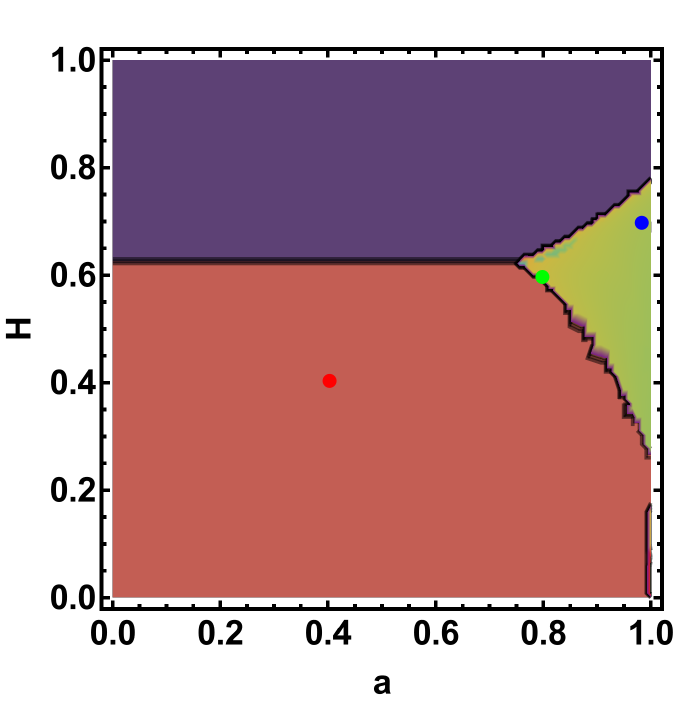}
\put(50,-3){\Large $\boldsymbol{b}$}
\put(-13,49){\Large $\boldsymbol{|H|}$}
\put(40,98){\Large $\boldsymbol{\phi = 60^\circ}$}
\put(44.5,44.5){$\boldsymbol{A}$}
\put(69,55){$\boldsymbol{B}$}
\put(84,60.5){$\boldsymbol{C}$}
\end{overpic}&
\begin{overpic}[width=0.48\linewidth,trim={20 20 0 0},clip]{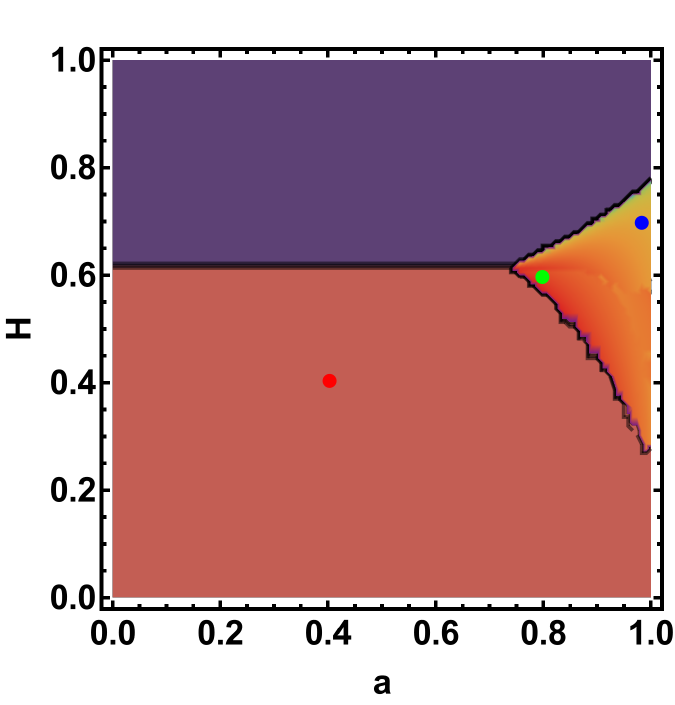}
\put(50,-3){\Large $\boldsymbol{b}$}
\put(40,98){\Large $\boldsymbol{\phi = 90^\circ}$}
\put(44.5,44.5){$\boldsymbol{A}$}
\put(69,54.2){$\boldsymbol{B}$}
\put(84,60.5){$\boldsymbol{C}$}
\end{overpic}
\end{tabular}
\caption{
Phase diagrams for the model with varying DMI parameter $b\in[0,1]$ and applied field strength $|\Hvec|$ for fields directed along 
$\hat\Hvec=(\sin\theta\cos\phi,\sin\theta\sin\phi,\cos\theta)$ with fixed angle of latitude $\theta=\pi/8$, but angles of latitude $\phi\in\{0,\pi/6,\pi/3,\pi/2\}$. 
The colouring conventions are the same as in \figref{fig:fixDMI}, and 
the energetically optimal solutions at the labelled points are shown in \figref{fig:fixedHconfigs2}. }
\label{fig:fixH2}
\end{figure}

\begin{figure}
\begin{tabular}{|c|ccc|}
\hline
{\Large \parbox[c][.8cm][c]{.8cm}{\centering\rotatebox{90}{$\qquad\qquad\qquad\;\boldsymbol{\phi=0^\circ}$}}} &
\begin{overpic}[width=0.25\linewidth]{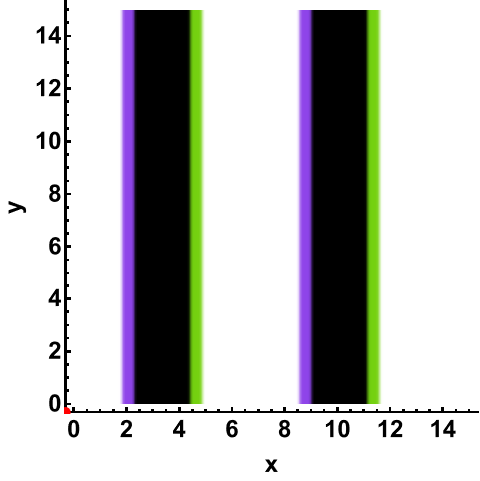}
\put(11.8,11.7){\tikzcircle[red]{2pt}}
\put(-3,84){$\boldsymbol{A}$}
\end{overpic}&
\begin{overpic}[width=0.25\linewidth]{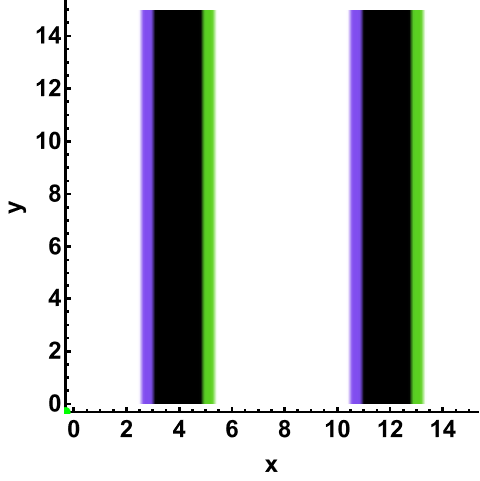}
\put(11.8,11.7){\tikzcircle[green]{2pt}}
\put(-3,84){$\boldsymbol{B}$}
\end{overpic}&
\begin{overpic}[width=0.25\linewidth]{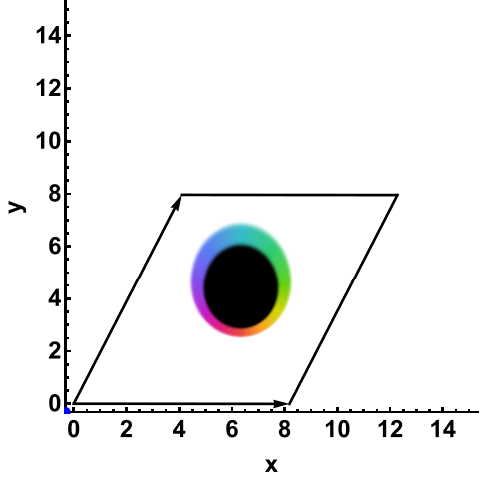}
\put(11.8,11.7){\tikzcircle[blue]{2pt}}
\put(-3,84){$\boldsymbol{C}$}
\end{overpic}\\
\hline
{\Large \parbox[c][.8cm][c]{.8cm}{\centering\rotatebox{90}{$\qquad\qquad\qquad\quad\boldsymbol{\phi=30^\circ}$}}}&
\begin{overpic}[width=0.25\linewidth]{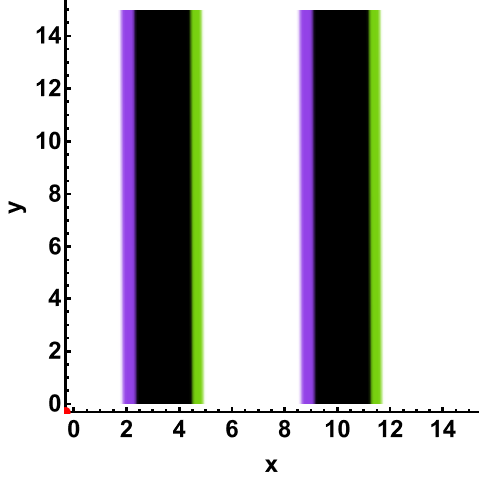}
\put(11.8,11.7){\tikzcircle[red]{2pt}}
\put(-3,84){$\boldsymbol{A}$}
\end{overpic}&
\begin{overpic}[width=0.25\linewidth]{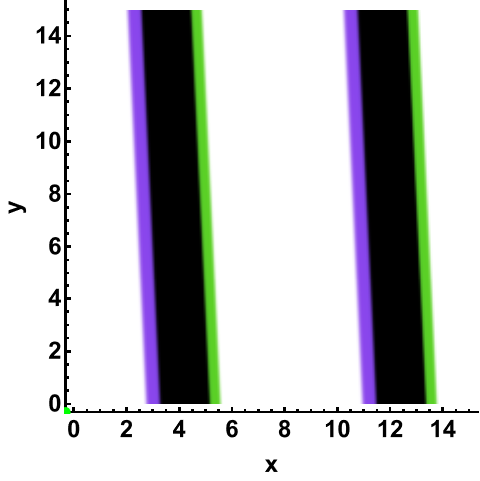}
\put(11.8,11.7){\tikzcircle[green]{2pt}}
\put(-3,84){$\boldsymbol{B}$}
\end{overpic}&
\begin{overpic}[width=0.25\linewidth]{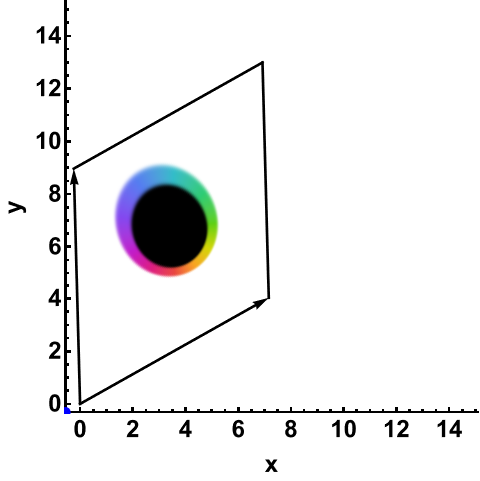}
\put(11.8,11.7){\tikzcircle[blue]{2pt}}
\put(-3,84){$\boldsymbol{C}$}
\end{overpic}\\
\hline
{\Large \parbox[c][.8cm][c]{.8cm}{\centering\rotatebox{90}{$\qquad\qquad\qquad\quad\boldsymbol{\phi=60^\circ}$}}}&
\begin{overpic}[width=0.25\linewidth]{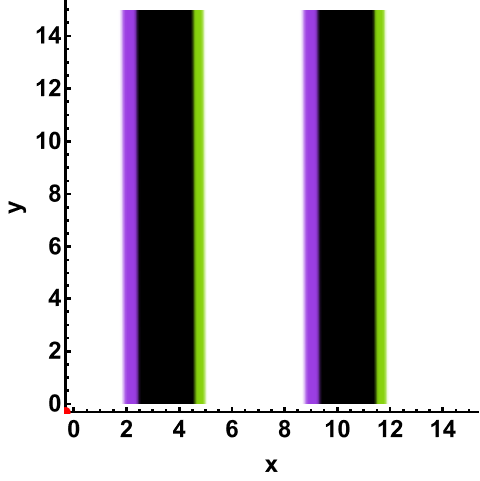}
\put(11.8,11.7){\tikzcircle[red]{2pt}}
\put(-3,84){$\boldsymbol{A}$}
\end{overpic}&
\begin{overpic}[width=0.25\linewidth]{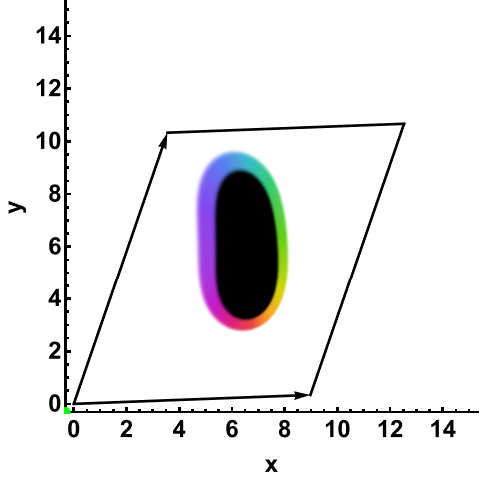}
\put(11.8,11.7){\tikzcircle[green]{2pt}}
\put(-3,84){$\boldsymbol{B}$}
\end{overpic}&
\begin{overpic}[width=0.25\linewidth]{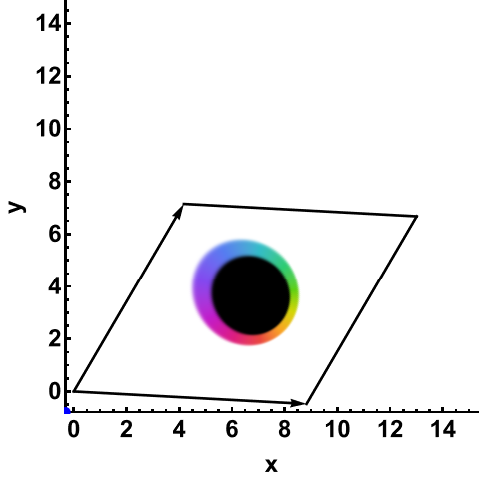}
\put(11.8,11.7){\tikzcircle[blue]{2pt}}
\put(-3,84){$\boldsymbol{C}$}
\end{overpic}\\
\hline
{\Large \parbox[c][.8cm][c]{.8cm}{\centering\rotatebox{90}{$\qquad\qquad\qquad\quad\boldsymbol{\phi=90^\circ}$}}}&
\begin{overpic}[width=0.25\linewidth]{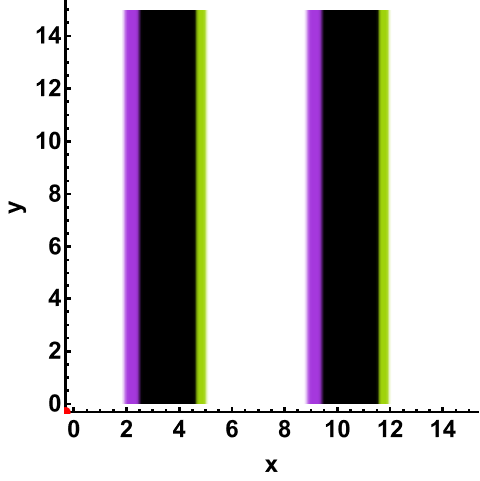}
\put(11.8,11.7){\tikzcircle[red]{2pt}}
\put(-3,84){$\boldsymbol{A}$}
\end{overpic}&
\begin{overpic}[width=0.25\linewidth]{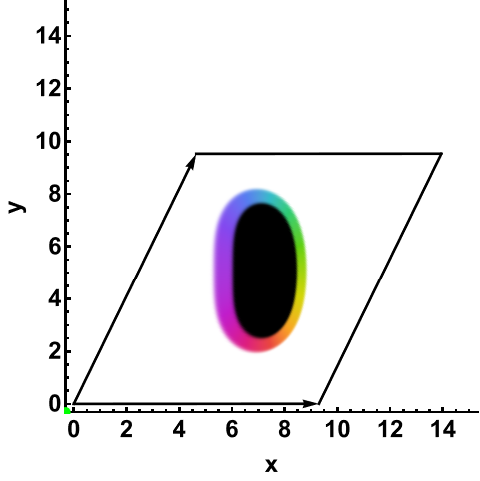}
\put(11.8,11.7){\tikzcircle[green]{2pt}}
\put(-3,84){$\boldsymbol{B}$}
\end{overpic}&
\begin{overpic}[width=0.25\linewidth]{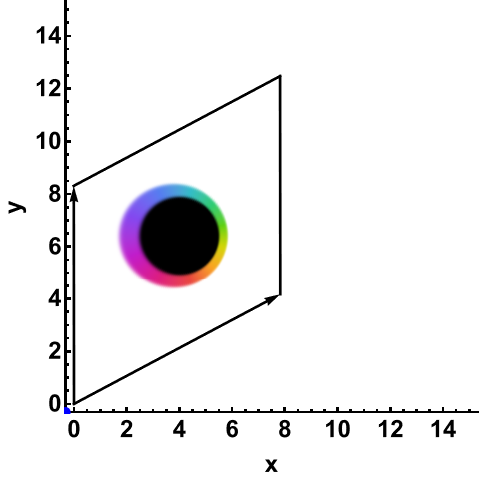}
\put(11.8,11.7){\tikzcircle[blue]{2pt}}
\put(-3,84){$\boldsymbol{C}$}
\end{overpic}\\
\hline
\end{tabular}
\caption{Three optimal configurations for applied field $\hat{\Hvec} = (\sin \theta \cos \phi, \sin \theta \sin \phi, \cos\theta)$ where the angle of latitude is  fixed as $\theta = \pi/8$ but $\phi\in\{0,\pi/6,\pi/3,\pi/2\}$. The corresponding DMI parameters $b$ and field strengths $|\Hvec|$ are with different DMI terms as marked on the corresponding phase diagrams in \figref{fig:fixH2}. For skyrmion lattices (B,C) the unit cell with period vectors is shown and the spiral state (A) has translation invariance. The colouring of the configurations is explained in \figref{fig:skyrmion}.}
\label{fig:fixedHconfigs2}
\end{figure}

\figref{fig:fixH2} shows $(b,|H|)$ phase diagrams for four applied field orientations \newline $\hat\Hvec=(\sin\theta,\cos\phi,\sin\theta\cos\phi,\cos\theta)$, all with angle of latitude fixed at $\theta=\frac\pi8$, but with azimuthal angle
$\phi\in\{0,\pi/6,\pi/3,\pi/2\}$. Once again the optimal fields at the marked points are displayed in \figref{fig:fixedHconfigs2}. Note that the skymrion lattices become strongly distorted as we approach the boundary of the SK domain.

\section{Conclusions and outlook}
In this paper,  we mapped out  the phase space of chiral magnets in the plane whose total Hamiltonian includes the standard Heisenberg contributions as well as the most general DMI and Zeeman terms. 
For each point in the four dimensional parameter space we used a combination of analytical and numerical methods to determine the ground state of the theory, defined as the configuration with the lowest energy per unit area. 

There are three phases, distinguished by their symmetry: the spatially homogeneous ferromagnetic vacuum with zero energy, spiral states with translation invariance in one direction and negative energy per unit area, and skyrmion lattices, also with negative energy per unit area. As expected, the ground state is  always ferromagnetic for sufficiently strong magnetic fields. For sufficiently weak magnetic fields, spiral states have the lowest energy per unit area for any value of the irreducible  DMI parameter $b$. However, their precise nature depends on the  value of $b$  and on the direction of the magnetic field relative to the DMI vectors in a way that we have only begun to explore in this paper. Finally, magnetic skyrmion phases appear for intermediate magnetic field strengths, but only for a small range of  values for $b$, and only for magnetic fields fairly close to orthogonal to the plane spanned by the DMI vectors. This includes  the axisymmetric $b=1$ case  which supports skyrmion phases for the largest range of magnetic field strengths and orientations.  

Our results are consistent with the extensive mathematical literature on the stability of chiral magnetic skyrmions. In the much studied case of an axisymmetric Bloch DMI term together with a Zeeman and anisotropy term, it is known that magnetic skyrmions minimize the energy in the plane among all configurations with degree $-1$, provided the strength of the  DMI term is bounded by that of the Zeeman term in a suitable sense \cite{Melcher2014chiral}, but that the skyrmion becomes unstable in a regime where the DMI term dominates   and where helical configurations can have arbitrarily negative energy \cite{Ibrahim2023_CMP}. Similar stability results hold for the model with an axisymmetric DMI term in compact domains, when the ferromagnetic vacuum is  imposed as a Dirichlet boundary condition. Magnetic skyrmions minimize the energy in such models, provided the DMI coefficient is sufficiently small   \cite{MonteilMuratovSimonSlastikov2023}. 

The picture that emerges from \cite{Melcher2014chiral,Ibrahim2023_CMP}  is that, for a fixed strength of the DMI terms,  skyrmions of degree $-1$ and  with negative energy  exist for finite and often small  range of  magnetic field strengths. There is evidence from the the study of asymptotic forces in \cite{barton2023stability} and the absence of degree $-2$ solutions, that the degree $-1$ skyrmions always repel.  As anticipated in our Introduction and confirmed in our analysis, the combination of negative energies per skyrmion and the repulsion between them allows for the formation of lattices. This is quite different from the more familiar mechanism for the formation of soliton lattices through attractive forces between solitons of positive energy. Intuitively, one can think of the chiral magnetic skyrmions lattices as a  dynamical equilibrium state, where skyrmions are constantly moving away from each other and new skyrmions form in the void between them to lower the total energy of the configuration.

It would clearly be interesting to  have unified and, if possible, rigorous analytical understanding of ground states in chiral magnets, combining and linking results on energy minimizers for fixed topological charge in the plane \cite{Melcher2014chiral, Ibrahim2023_CMP} or in compact domains \cite{MonteilMuratovSimonSlastikov2023} with ours on energy minimizers in the plane for any topological charge.  Since our method for finding skyrmion lattices relies on finding energy minimizing configurations for  a fixed torus,  a first and  important step would be to understand when there are negative energy skyrmions on a fixed torus, for arbitrary DMI terms and external magnetic fields. 

\subsection*{Acknowledgements}
This work was funded by the UK Engineering and Physical Sciences research council through grant EP/Y033256/1, and benefited from two of the authors (Speight and Winyard) participating in the COST action CA23134 (Topological textures in condensed matter).

\bibliographystyle{unsrt}
\bibliography{bibliography}

\end{document}